\documentclass[11pt, a4paper]{article}
\usepackage{jcappub}

\usepackage{bbm}
\usepackage{slashed}
\usepackage{color}

\usepackage{epsfig,graphicx,bm,color} % Include figure files
\usepackage{dcolumn}% Align table columns on decimal point
\usepackage{bm}% bold math
\usepackage{hyperref}% add hypertext capabilities
\usepackage{color}
\usepackage{breakurl}
\usepackage{soul} 
\newcommand{\be}{\begin{equation}}
\newcommand{\ee}{\end{equation}}
\newcommand{\bea}{\begin{eqnarray}}
\newcommand{\eea}{\end{eqnarray}}

\newcommand{\lsim}{\mathrel{\mathop{\kern 0pt \rlap
  {\raise.2ex\hbox{$<$}}}
  \lower.9ex\hbox{\kern-.190em $\sim$}}}
\newcommand{\gsim}{\mathrel{\mathop{\kern 0pt \rlap
  {\raise.2ex\hbox{$>$}}}
  \lower.9ex\hbox{\kern-.190em $\sim$}}}

\newcommand{\Fermi}{{\it Fermi}}
\newcommand{\pMSSM}{pMSSM}
\newcommand{\sgn}[1]{\text{sgn}(#1)}

\newcommand{\likeJ}{\mathcal{L}_{\rm Joint}}
\newcommand{\like}{\mathcal{L}}
\newcommand{\Ohsq}{\Omega_\chi h^2}

\newcommand{\BR}{BR}
\newcommand\RBtaunu{\frac{\BR(B_u \to \tau \nu)}{\BR(B_u \to \tau \nu)_{SM}}}
\newcommand\DeltaO{\Delta_{0-}}

\newcommand\Rl{R_{l23}}

\newcommand\Dstaunu{\BR(D_s \to \tau \nu)}

\newcommand\brbsmumu{\BR(\overline{B}_s\to\mu^+\mu^-)}
\newcommand\brbdmumu{\BR(\overline{B}_d\to\mu^+\mu^-)}

\newcommand{\brbsgamma}{BR(\bar{B} \rightarrow X_s\gamma) }

\newcommand{\cl}{\text{CL}}

\begin{document}

% title to be defined
\title{Global analysis of the pMSSM in light of the Fermi GeV excess: prospects
for the LHC Run-II and astroparticle experiments}
%\title{Global fits of the pMSSM-19 in light of the Fermi GeV excess: prospects for run 2 of the LHC}
%\title{A global analysis of the pMSSM: From the Galactic center GeV excess to LHC run II}

\author[a]{Gianfranco Bertone,}
%\emailAdd{g.bertone@uva.nl}
\author[a]{Francesca Calore,}
%\emailAdd{f.calore@uva.nl}
\author[b]{Sascha Caron,}
%\emailAdd{}
\author[c]{Roberto Ruiz,}
%\emailAdd{}
\author[d]{Jong Soo Kim,}
%\emailAdd{}
\author[e]{Roberto Trotta,}
\author[a]{Christoph Weniger.}

\affiliation[a]{GRAPPA, University of Amsterdam, Science Park 904,
  1090 GL Amsterdam, Netherlands} 
\affiliation[b]{IMAPP, Radboud University Nijmegen, P.O. Box 9010, NL-6500 GL Nijmegen, The Netherlands and
                Nikhef, Science Park, Amsterdam, The Netherlands}
\affiliation[c]{Instituto de F\'isica Corpuscular, IFIC-UV/CSIC, Valencia, Spain}
\affiliation[d]{Instituto de F\'{\i}sica Te\'{o}rica UAM/CSIC, Madrid, Spain}
\affiliation[e]{Imperial Centre for Inference and Cosmology, Imperial College London, Blackett Laboratory, Prince Consort Road, London SW7 2AZ, UK}

\abstract{
We present a new global fit of the 19-dimensional phenomenological Minimal Supersymmetric Standard Model (pMSSM-19) that comply with all the latest experimental results from dark matter indirect, direct and accelerator dark matter searches. We show that the model provides a satisfactory explanation of the excess of gamma-rays from the Galactic centre observed by the \Fermi~Large Area Telescope, assuming that it is produced by the annihilation of neutralinos in the Milky Way halo. We identify two regions that pass all the constraints: the first corresponds to neutralinos with a mass $\sim 80-100$ GeV annihilating into $WW$ with a branching ratio of 95\% ; the second to heavier neutralinos, with mass $\sim 180-200$ GeV annihilating into $\bar{t}t$ with a branching ratio of 87\%. 
We show that neutralinos compatible with the Galactic centre GeV excess will soon be within the reach of LHC run-II -- notably through searches for 
charginos and neutralinos, squarks and light smuons -- and of Xenon1T, thanks to its unprecedented sensitivity to spin-dependent cross-section off neutrons.
}

\maketitle

%%%%%%%%%%%%%%%%%%%%%%%%%%%%%%%%%%%%%%%%
%%%%%%%%%%%%%%%%%%%%%%%%%%%%%%%%%%%%%%%%

\section{Introduction}
\label{sec:intro}
%The dark matter quest. --
The existence of dark matter (DM) is now robustly established \cite{Jungman96,Bergstrom00,Bertone05,BertoneBook} and its cosmological abundance measured with high precision~\cite{Planck:2006uk}.
Yet, the fundamental nature of the most abundant matter component in the Universe is unknown.
According to the most promising theories, DM is a \emph{new fundamental particle}. As a consequence,
the search for DM is also a search for new physics beyond that of the well-known elementary particles, 
as laid down in the Standard Model (SM). 
Weakly interacting massive particles (WIMPs) are the leading DM candidates: 
they arise in many extensions of the Standard Model of particle physics, and naturally achieve the appropriate relic density through self-annihilation the early Universe. 
WIMPs can be searched for with three detection strategies: direct detection 
of the energy recoil of nuclei scattering off DM particles; indirect detection of the final stable
products of DM annihilation or decay, as for example gamma rays; and accelerator searches for new particles, in particular at the Large Hadron Collider (LHC).

\smallskip

%The \Fermi~GeV excess. -- 
An excess in the $\gamma$-ray emission from the centre of our Galaxy 
has been discovered in data from the Large Area telescope (LAT), aboard the \Fermi~satellite~\cite{Goodenough:2009gk, 
Vitale:2009hr, Hooper:2010mq, Hooper:2011ti, Abazajian:2012pn,
Gordon:2013vta, Hooper:2013rwa, Abazajian:2014fta, Daylan:2014rsa,
Calore:2014xka, fermigc}.
The nature of the so-called \Fermi~GeV excess remains a mystery. Several explanations have been put forward, the most exciting of which is perhaps DM annihilation in the halo of the Milky Way (see for example~\cite{Calore:2014nla,Daylan:2014rsa,Abazajian:2014fta}). 
Astrophysical
processes have also been suggested: the emission from a population of dim unresolved
sources~\cite{Abazajian:2010zy,
Calore:2014oga, Yuan:2014rca, Petrovic:2014xra}, and the inverse Compton emission from a new population of 
cosmic-ray, either from 
time-dependent events taking place in the past of the Galaxy~\cite{Carlson:2014cwa,
Petrovic:2014uda,  Cholis:2015dea} or from the high star formation
activity in the inner Galaxy~\cite{Gaggero:2015nsa}.

Very recently, two reanalyses of the gamma-ray emission from the inner Galaxy
found strong evidence for the \Fermi~GeV excess being due to
hundreds to thousands of dim or unresolved point sources~\cite{Bartels:2015aea,
Lee:2015fea}, most likely millisecond pulsars~\cite{Abazajian:2010zy}.  Even
more recently, it was shown that part or all of
the required millisecond pulsar population could originate from the disruption of
globular clusters by tidal forces in the inner Galaxy~\cite{Brandt:2015ula}.
Directly detecting some of the millisecond pulsars in the inner Galaxy by radio
observations is the next critical step for fully establishing this
scenario~\cite{Bartels:2015aea}. 

\medskip

It is however striking that the \Fermi~GeV excess spectrum and spatial distribution are well fitted by what is expected
from DM annihilation. The excess could be the first non-gravitational signal of DM particles.   
It is thus urgent to either corroborate or disprove the DM particle nature of the \Fermi~GeV excess
in the framework of concrete models for physics beyond the SM.
Supersymmetry (SUSY) is one of the best-motivated classes of renormalizable extensions of the SM, 
which can accommodate a stable DM particle together with new degrees of freedom that mediate interactions. 
The most generic R-parity conserving and phenomenologically viable supersymmetric model is the
 phenomenological Minimal Supersymmetric Standard Model (pMSSM)~\cite{Djouadi:1998di}. 
In this work we address the following key question: can models of the pMSSM explain the observed properties
of the \Fermi~GeV excess, while retaining consistency with other experimental data? And if so, what are the detection prospects for future direct detection and collider 
experiments?

Until now, MSSM scenarios~\cite{Cahill-Rowley:2014ora} could not reproduce the \Fermi~GeV excess as observed by~\cite{Daylan:2014rsa}, as it was impossible to obtain in this 
framework a light neutralino ($m_{\chi} \sim$ 30 -- 40 GeV), as required to fit the  \Fermi~GeV excess spectrum, which could also account for the cosmological DM as measured by Planck~\cite{Planck:2006uk}.
However, \cite{Calore:2014xka} demonstrated that higher WIMP masses and
annihilation channels different from b-quark pairs can give a good fit to the 
\Fermi\ excess, owing to the freedom allowed by background model systematics.
By fitting the GeV excess data of~\cite{Calore:2014xka}, it has been shown that viable solutions in the 
MSSM exist~\cite{Agrawal:2014oha}.
In the context of the pMSSM, \cite{Caron:2015wda} demonstrated that a re-assessment of the theoretical uncertainties in the DM signal spectra opens up a new phenomenology at the LHC experiments.
Here we present the \emph{first} systematic study of the pMSSM parameter space through global fitting techniques.
This approach exhaustively covers all possible phenomenological signatures, allowing us a complete overview of the viable pMSSM interpretation
of the \Fermi~GeV excess. 
%Our goal --

\bigskip

The paper is organized as follows: in section~\ref{sec:theory} we briefly present the well-known framework of the pMSSM
and the parameters describing the model. In section~\ref{sec:setup} we describe the experimental set-up of the global fit
and the implementation of the joint likelihood. We present the results of the parameter scan in section~\ref{sec:results}.
We then discuss the implications for future direct and indirect detection experiments in section~\ref{sec:implicationsDDID}
and the prospects for detection at the LHC run II in section~\ref{sec:LHCII}.
Finally, we summarize our conclusions in section~\ref{sec:conclusions}.

%%%%%%%%%%%%%%%%%%%%%%%%%%%%%%%%%%%%%%%%
%%%%%%%%%%%%%%%%%%%%%%%%%%%%%%%%%%%%%%%%
\section{The theoretical framework: the pMSSM}
\label{sec:theory}
We here study the pMSSM~\cite{Djouadi:1998di}, in which
the number of free parameters can be reduced to 19, given the present lack of experimental evidence for SUSY 
%Indeed, while highly constrained models as the constrained-MSSM are under pressure in the light of the recent negative
%sparticle searches at the LHC, there is 
and no experimental indication that one requires the full freedom 
of a 22-dimensional pMSSM at present. 

In this model, the lightest supersymmetric particle is the lightest neutralino, $\chi$, a combination the
neutral electroweak gauginos and higgsinos fields. The neutralino is one of the most well-motivated particle
DM candidates since it is neutral, stable over cosmological timescales and can lead naturally to the correct DM relic abundance
in the early Universe. In what follow, we assume that the neutralino is the particle DM candidate although we do 
not impose it to fully account for the DM relic abundance as measured by Planck but we allow 
for subdominant contributions to the DM content (cf.~section~\ref{sec:setup}).

We assume first and second generation mass universality, separately in the lepton 
and quark sectors (table~\ref{pMSSM_priors}). 
The trilinear couplings of the sfermions enter in the off-diagonal parts of the sfermion mass matrices. 
Since these entries are proportional to the Yukawa couplings of the 
respective fermions, we approximate the trilinear couplings associated 
with the first and second generation fermions to be zero, while the 
parameters $A_t$,  $A_b$ and $A_\tau$ represent the third generation 
trilinear couplings.
In our set-up, the Higgs sector is fully described by the ratio of the Higgs vacuum expectation values $\tan \beta$,
the higgsino mass parameter $\mu$ and the mass of the pseudoscalar Higgs $m_A$, which
are more directly related to the phenomenology of the model.
This 19-dimensional realization of the pMSSM encapsulates all phenomenologically relevant features of the full model that are of interest for DM and collider experiments. The model parameters are displayed in table~\ref{pMSSM_priors}, along with their prior ranges. All of the input parameters are defined at the SUSY scale $\sqrt{m_{\tilde{t_1} } m_{\tilde{t_2} } }$.

\begin{table}
\begin{center}
\begin{tabular}{l l | l l}
\hline
\hline
\multicolumn{4}{c}{\pMSSM\ parameters and priors} \\
\multicolumn{2}{c}{Flat priors} & \multicolumn{2}{c}{Log priors} \\
\hline
$M_1$ [TeV] & (-5, 5) & $\sgn{M_1} \log |M_1|/\text{GeV}$ & $(-3.7, 3.7)$ \\
$M_2 $ [TeV] & (0.1, 5) & $\log M_2/\text{GeV}$ & $(2,3.7)$ \\
$M_3 $ [TeV] & (-5, 5) & $\sgn{M_3} \log |M_3|/\text{GeV}$ & $(-3.7, 3.7)$ \\
$m_{L} $ [TeV] & (0.1,4) & $\log m_L/\text{GeV}$ & $(-1,3.6)$ \\
$m_{E} $ [TeV] & (0.1,4) & $\log m_E/\text{GeV}$ & $(-1,3.6)$ \\
$m_{L_3} $ [TeV] & (0.1,4) & $\log m_{L_3}/\text{GeV}$ & $(-1,3.6)$ \\
$m_{E_3} $ [TeV] & (0.1,4) & $\log m_{E_3}/\text{GeV}$ & $(-1,3.6)$ \\
$m_{Q} $ [TeV] & (0.1,4) & $\log m_{Q}/\text{GeV}$ & $(-1,3.6)$ \\
$m_{U} $ [TeV] & (0.1,4) & $\log m_{U}/\text{GeV}$ & $(-1,3.6)$ \\
$m_{D} $ [TeV] & (0.1,4) & $\log m_{D}/\text{GeV}$ & $(-1,3.6)$ \\
$m_{Q_3} $ [TeV] & (0.1,4) &$\log m_{Q_3}/\text{GeV}$ & $(-1,3.6)$\\
$m_{U_3} $ [TeV] & (0.1,4) & $\log m_{U_3}/\text{GeV}$ & $(-1,3.6)$ \\
$m_{D_3} $ [TeV] & (0.1,4) & $\log m_{D_3}/\text{GeV}$ & $(-1,3.6)$ \\
$A_t $ [TeV] & (-10, 10) & $\sgn{A_t} \log |A_t|/\text{GeV}$ & $(-4, 4)$  \\
$A_b $ [TeV] & (-10,10) & $\sgn{A_0} \log |A_b|/\text{GeV}$ & $(-4, 4)$ \\
$A_\tau $ [TeV] & (-10,10) & $\sgn{A_0} \log |A_\tau|/\text{GeV}$ & $(-4, 4)$ \\
$\mu $ [TeV] & (-5,5) & $\sgn{\mu} \log |\mu|/\text{GeV}$ & $(-3.7, 3.7)$ \\
$m_A $ [TeV] & (0.01, 5) & $\log m_A/\text{GeV}$ & $(1, 3.7)$ \\
$\tan\beta$ & $(2, 62)$ & $\tan\beta$ & $(2, 62)$ \\
\hline
%& Gaussian prior  & Range scanned (170.6, 175.8) \\
$M_t$ [GeV] & \multicolumn{3}{c}{$173.2 \pm 0.87$~\cite{ATLAS:2014wva} (Gaussian prior)}   \\
$\rho_0$ [GeV/cm$^3$] & \multicolumn{3}{c}{$0.4 \pm 0.1$~\cite{Pato:2010zk} (Gaussian prior)}   \\
\hline
\end{tabular}
\end{center}
\caption{pMSSM parameters and top mass value used in this paper and the 
prior range for the two prior choices adopted in our scans. ``Flat priors'' are 
uniform on the parameter itself (within the ranges indicated), while ``Log priors'' are uniform in the log 
of the parameter (within the ranges indicated).}\label{pMSSM_priors}
\end{table}

%%%%%%%%%%%%%%%%%%%%%%%%%%%%%%%%%%%%%%%%
%%%%%%%%%%%%%%%%%%%%%%%%%%%%%%%%%%%%%%%%
\section{The experimental setup}
\label{sec:setup}

We implement experimental constraints with a joint likelihood function,  whose logarithm takes the following form:
\begin{equation}
\begin{aligned}
\ln \likeJ &=  \ln \like_{\rm GCE} +  \ln \like_{\rm EW}+ \ln \like_{\rm B(D)} + \ln \like_{\Ohsq} \\
 &+ \ln \like_{\rm LUX} + \ln \like_{\rm IC}  +  \ln \like_{\rm Higgs} + \ln \like_{\rm SUSY},     
\end{aligned}
\label{eq:like}
\end{equation}
where $\like_\text{GCE}$ represents the \Fermi~GeV excess,  $\like_\text{EW}$ electroweak precision observables, 
$\like_\text{B(D)}$ B and D physics constraints,  $\like_{\Ohsq}$ measurements of the cosmological DM relic density, $\like_\text{LUX}$ ($\like_\text{IC}$)
direct (indirect) DM detection constraints and $\like_\text{Higgs}$ ($\like_\text{SUSY}$) Higgs (sparticles) searches at colliders. 
We discuss each component in turn:

\paragraph{$\like_{\rm GCE}$:} For the \Fermi~GeV excess likelihood we follow the
treatment in~\cite{Calore:2014xka}, to account astrophysical uncertainties.  In
addition, we include uncorrelated 10\% uncertainties as DM modeling
systematics, following ref.~\cite{Caron:2015wda}.  We marginalize the
likelihood function over the uncertainties in the Galactic centre $J$-value, which is assumed to follow a log-normal distribution with mean $
\log_{10} J/(\rm GeV^2\,cm^{-5}\,sr^{-1}) = 23.29$ and variance $\Delta\log_{10} J/
(\rm GeV^2\,cm^{-5}\,sr^{-1}) = 0.37$~\cite{Calore:2014nla}.
When the predicted neutralino relic density, $\Omega_\chi$, is smaller than the Planck measurement, 
$\Omega_{\rm DM}$, we follow~\cite{Bertone:2010rv} and adopt the so-called ``scaling Ansatz'', i.e. we assume that the local ratio of neutralino ($\rho_\chi$) to total DM densities
 ($\rho_{\rm DM}$) is equal to that for their cosmic abundances: 
\be \label{eq:scaling}
\rho_\chi / \rho_{\rm DM}= \Omega_\chi / \Omega_{\rm DM} \equiv f_\chi. 
\ee

\paragraph{$\like_{\rm EW}$:}This implements constraints from Z-pole 
measurements at LEP~\cite{ALEPH:2005ab}. We include the constraint on the effective electroweak mixing angle for leptons $\sin^2\theta_\text{eff}$, the total 
width of the Z-boson $\Gamma_{Z}$, the hadronic pole cross-section $\sigma^0_{had}$, as well as the decay width ratios $R^0_l$, $R^0_c$ and the asymmetry parameters 
$A_l$, $A_b$, $A_c$ and $A^{0,l}_{FB}$, $A^{0,c}_{FB}$. In addition, we also use the measurement of the mass of the W boson $m_W$ from the LEP experiment \cite{ALEPH:2005ab}. 
We apply a Gaussian likelihood for all of these quantities, with mean and standard deviation as given in table II of \cite{Strege:2014ija}. 

\paragraph{$\like_{\rm B(D)}$:} The flavor observables related with B and D physics considered are $\brbsgamma$, $R_{\Delta M_{B_s}}$, $\RBtaunu$,  $\DeltaO$, $\Dstaunu$, $\Rl$, 
$\brbsmumu$ and $\brbdmumu$. For all we apply a Gaussian likelihood and for most of them we use the measurements following table II of \cite{Strege:2014ija}. The experimental values 
assumed for $\brbsmumu$ and $\brbdmumu$ are $(2.9 \pm 0.8) \times 10^{-9}$ and $(3.6 \pm 1.55) \times 10^{-10}$ (theoretical uncertainties included \cite{Arbey:2012ax}) \cite{CMSandLHCbCollaborations:2013pla} respectively.

\paragraph{$\like_{\Ohsq}$:} We include the Planck Cosmic Microwave Background data constraint on the DM relic abundance as an upper limit,  to allow for the possibility that neutralinos 
are a sub-dominant DM component. We follow the formalism in the Appendix of~\cite{Bertone:2010ww}, using as central value the result from Planck temperature and lensing 
data $\Ohsq = 0.1186 \pm 0.0031$ \cite{Ade:2013zuv} with a (fixed) theoretical uncertainty, $\tau = 0.012$, to account for the numerical uncertainties entering in the calculation 
of the relic density.

\paragraph{$\like_{\rm LUX}$:} For DM direct detection we include upper limits from the LUX experiment \cite{Akerib:2013tjd}, as implemented in the LUXCalc 
code~\cite{Savage:2015xta}, including both the spin-independent and spin-dependent cross-sections in the event rate calculation. We adopt  
hadronic matrix elements determined by lattice QCD \cite{QCDSF:2011aa,Junnarkar:2013ac}. 

\paragraph{$\like_{\rm IC}$:} This implements conservative upper limits on the proton spin-dependent cross-section from the IceCube detector in its 79-string configuration \cite{Aartsen:2012kia} (IC-79). Comparable -- if slightly weaker -- limits have been set for the $WW$ channel by Super-Kamiokande~\cite{PhysRevLett.114.141301}
and ANTARES~\cite{Adrian-Martinez:2013ayv}.
The most stringent constraint is for the case where WIMPs annihilate exclusively to $WW$ pairs. Since the neutrino spectrum generated by $Z$ bosons is similar to the $W$ bosons 
we apply this constraint whenever the combined branching fraction to $WW$ and $ZZ$ is above 95\%. In that case the likelihood is a step function smeared with half a Gaussian (as in Eq.~(3.5) of \cite{deAustri:2006pe}) to account for theoretical and experimental uncertainties that we set to be 50\% of the predicted value.  

\paragraph{$\like_{\rm Higgs}$:} The likelihood for the Higgs searches has two components: the first implements bounds obtained from Higgs searches at LEP,  Tevatron and LHC via 
HiggsBounds \cite{Bechtle:2013wla}, which returns whether a model is excluded or not at the 95\% \cl. 
The second component constrains the mass and the production times decay rates of the Higgs-like boson discovered by the LHC experiments 
ATLAS \cite{Aad:2012tfa} and CMS \cite{Chatrchyan:2012ufa}. We use HiggsSignals \cite{Bechtle:2013xfa} assuming a theoretical uncertainty in the lightest Higgs mass calculation of 2 GeV.

\paragraph{$\like_{\rm SUSY}$:} SUSY searches constraints at LEP and Tevatron follow the likelihood used~\cite{deAustri:2006pe}. 
We have imposed the strict constraints from a large number of searches for supersymmetry at the LHC. The branching ratios of the sparticles have been calculated with  SUSYHIT 1.5 \cite{Djouadi:2006bz}. We have generated the hadronized event samples with Pythia 8.2 \cite{Sjostrand:2014zea} and have employed the NNPDF 2.3 parton distribution functions 
\cite{Ball:2012cx}. The generated events are passed on to CheckMATE \cite{Drees:2013wra}, which is based on the fast detector simulation Delphes 3.10 \cite{deFavereau:2013fsa}. 
CheckMATE tests if the model point in question is excluded or not at $95\%$ confidence level by comparing to current experimental searches at the LHC for supersymmetry in the 
relevant hadronic and leptonic final states with large missing transverse momentum. We assign a 0 log-likelihood $\like_{\rm SUSY} = 0$ if the point passes all constraints, 
and exclude it if it fails any of them.
 
We have only included observables we consider robust in order to be conservative. For instance, we have  dropped the 
electroweak precision observables $R^0_b$ and $A^{0,b}_{FB}$ in the fit because it is unclear whether the large deviations of 2.5$\sigma$ that are observed with respect 
to the SM predictions are due to unknown systematic uncertainties or to new physics. The experimental status of the magnetic anomaly of the muon, 
$a_\mu= \frac{1}{2}(g-2)_\mu$ remains unclear in the face of persistent discrepancies in the determination of the hadronic vacuum-polarization diagram 
using  either $e^+ e^-$ or the hadronic $\tau-$decay data.
We do not include DM searches in dwarf spheroidal galaxies by
the \Fermi-LAT~\cite{Ackermann:2015zua}, as constraints are given only for 100\% branching ratio into final states.

Finally, we refer to \cite{Strege:2014ija} for details about how the SUSY spectrum and observables are computed.

\begin{figure*}[t!]
\centering
\includegraphics[trim=4cm 3.5cm 5.5cm 6cm, clip=true, scale=0.6]{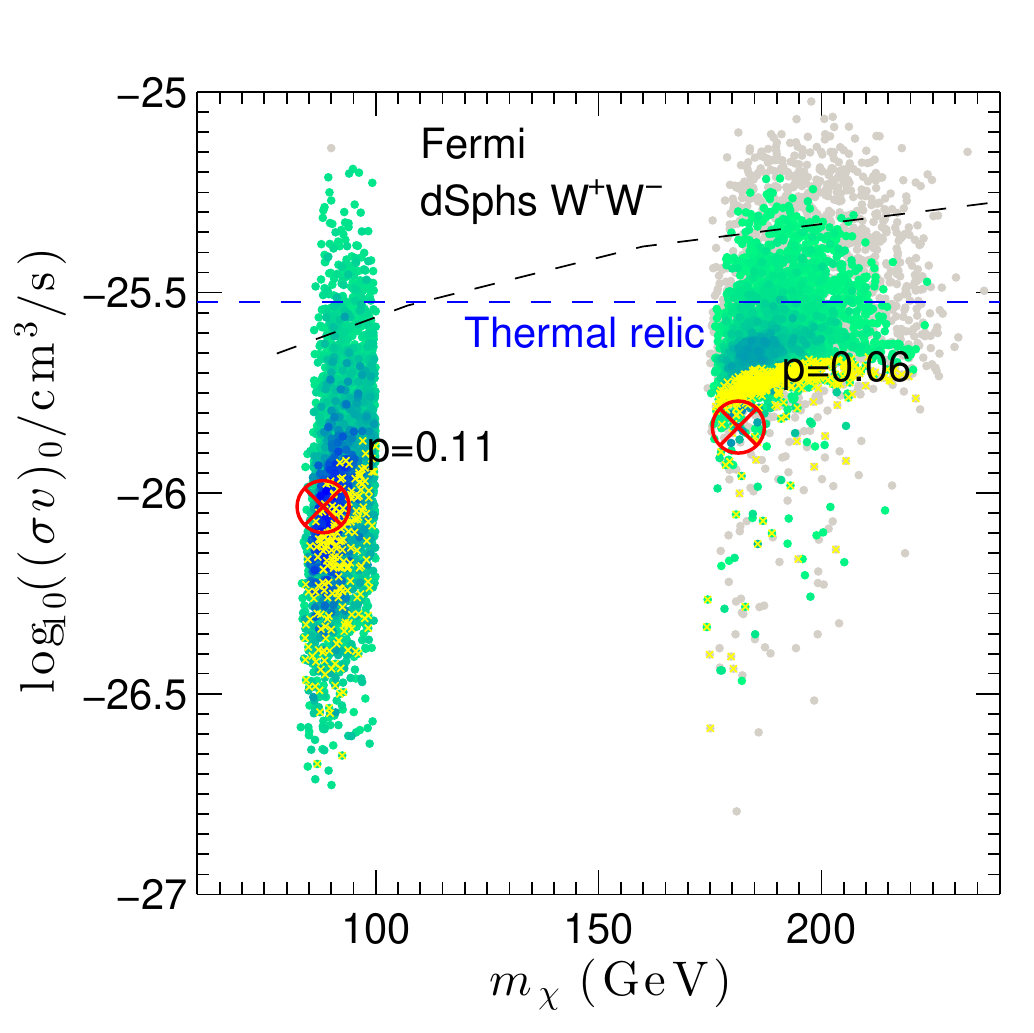} 
\includegraphics[trim=3.6cm 3.5cm 3cm 6cm, clip=true, scale=0.6]{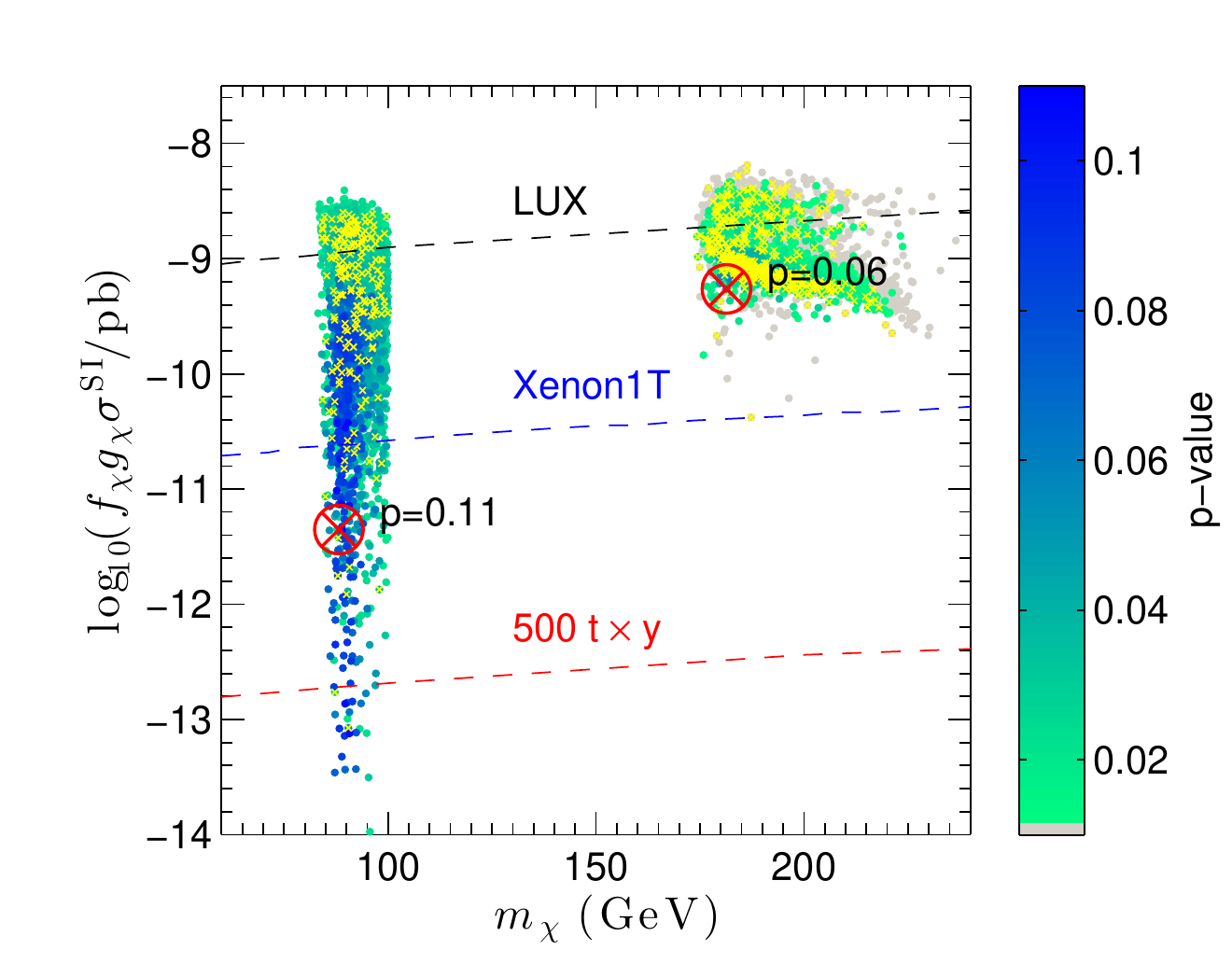} \\ 
\includegraphics[trim=4cm 3.5cm 5.5cm 6cm, clip=true, scale=0.6]{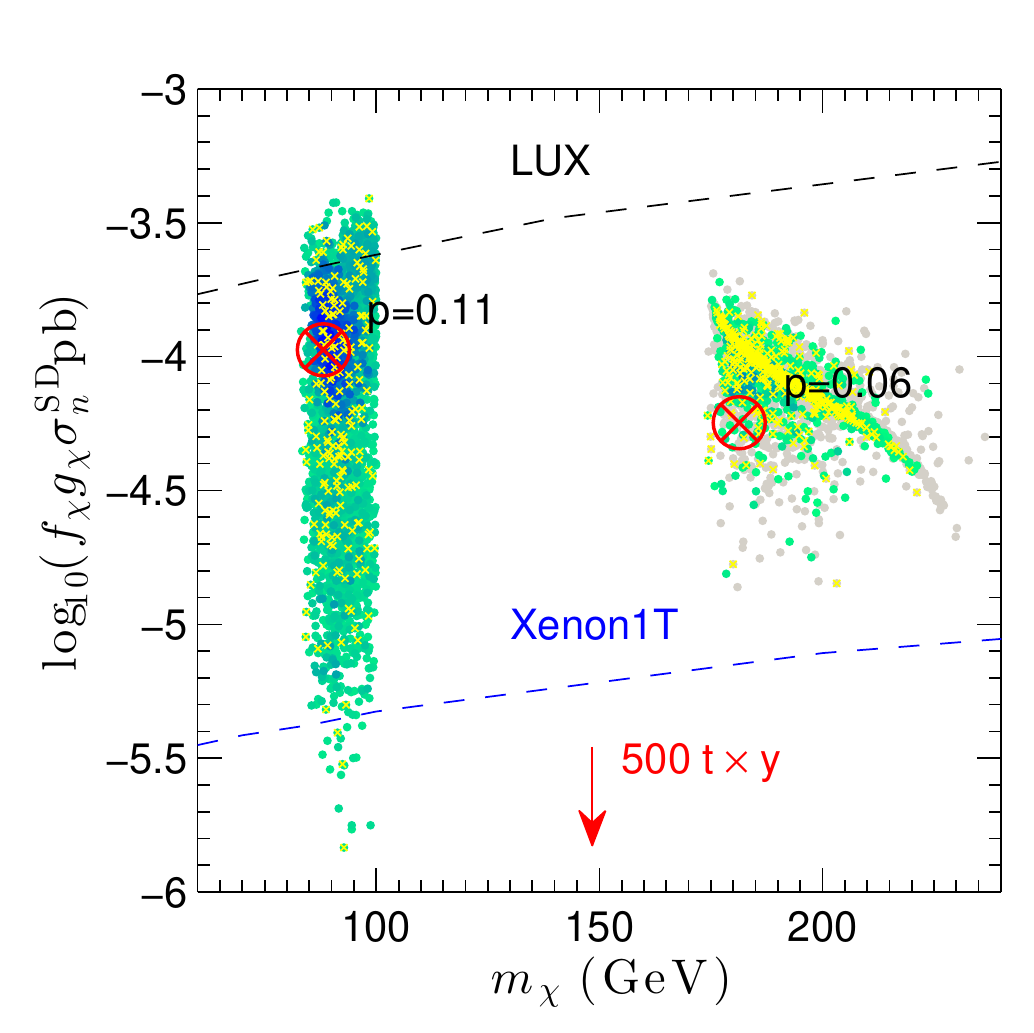}
\includegraphics[trim=3.6cm 3.5cm 3cm 6cm, clip=true, scale=0.6]{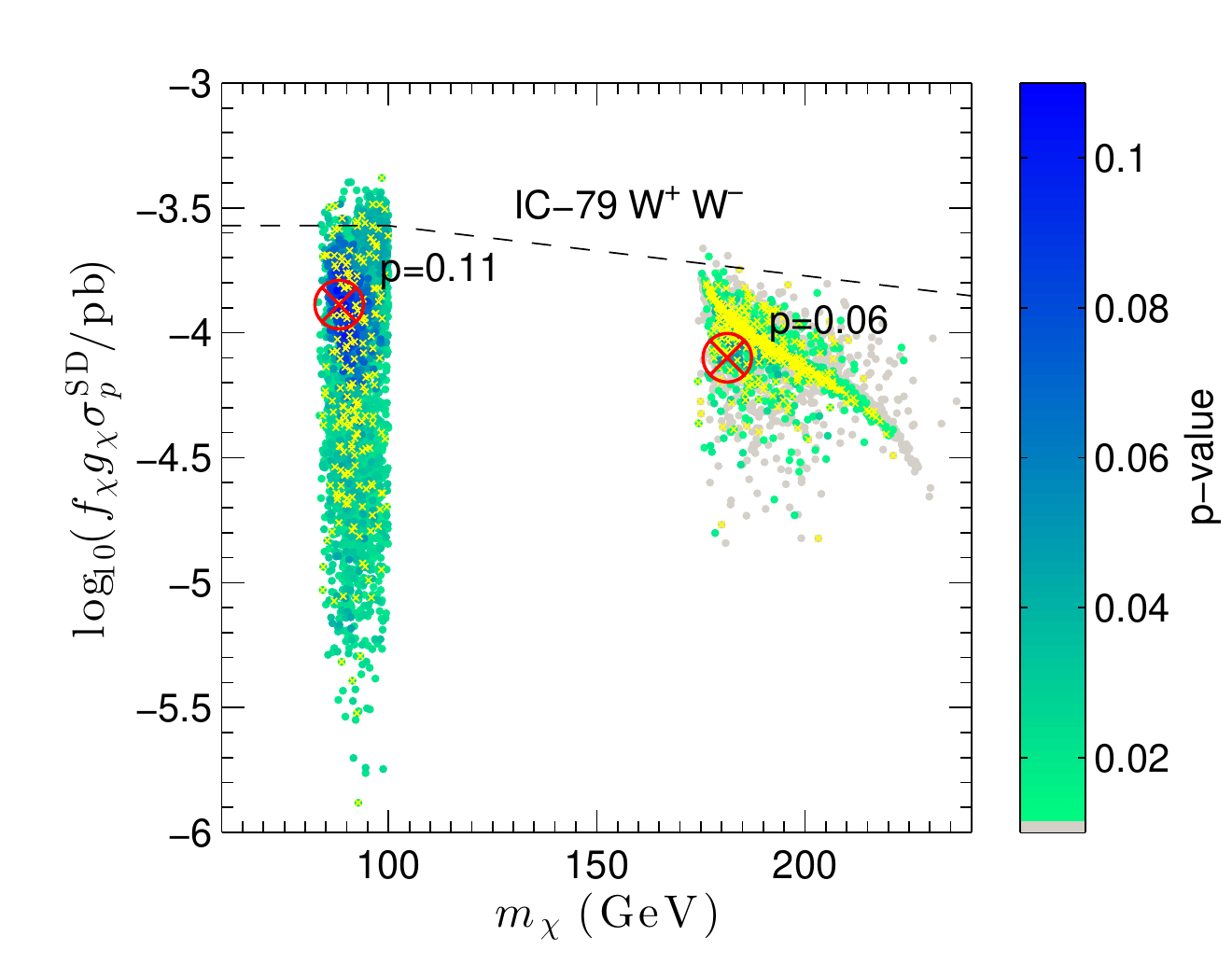}
%\includegraphics[trim=3cm 3cm 2cm 6cm, clip=true, scale=0.5]{figs/pvals-sigmav_mchi_fchi.pdf} % left, bottom, right and top \\
 % left, bottom, right and top \\
%\includegraphics[trim=8.5cm 15.5cm 6.5cm 25.5cm, clip=true, scale=0.22]{figs/pvals-sigmaSI_Oh2.jpg}
%\includegraphics[trim=8.5cm 15.5cm 6.5cm 25.5cm, clip=true, scale=0.22]{figs/pvals-mcharg-mchi.jpg}
%\includegraphics[trim=8.5cm 15.5cm 6.5cm 25.5cm, clip=true, scale=0.22]{figs/pvals-gluino_sbot.jpg} \\
\caption{2D map of the $p$-values of our fit, showing the annihilation (top left), the spin-independent (top right) and the spin-dependent (bottom panel, left for neutrons scattering and right for protons) cross-sections 
vs neutralino mass. The color-bar represents the $p$-value from the global fit. 
The yellow overlay highlights points that are within 2$\sigma$ of the \emph{Planck} relic density. Red crosses indicate the 
best-fit points in the two islands. The \Fermi\ dwarfs limit
for the $W^+W^-$ channel~\cite{Ackermann:2015zua} is plotted for reference
only, and it has not been applied. To compare with \Fermi\ dwarfs data, the
annihilation cross-section needs to be rescaled by a factor $f_\chi^2$, which
would suppress the signal well below current limits.  The spin-dependent and spin-independent cross-sections have
been multiplied by $f_\chi = \Omega_\chi/\Omega_\text{DM}$ and $g_\chi =
\rho_0/0.3$ GeV/cm$^3$ to facilitate comparison with current and future limits
(LUX~\cite{Akerib:2013tjd}, Xenon1T and a multi-ton liquid Xe detector with 500
t$\times$ yr exposure~\cite{2015arXiv150608309S}).   In the bottom right
panel, we display
the IC-79 limit~\cite{Aartsen:2012kia} used in our analysis. }
\label{fig:results}
\end{figure*}

We use the MultiNest~\cite{Feroz:2008xx} algorithm as implemented in SuperBayeS-v2.0, to perform a global fit of the pMSSM parameter space, including all the data in eq.~\eqref{eq:like}, excepting the SUSY searches at the LHC.
This is because the LHC searches evaluation is computationally too expensive to be performed on-the-fly. Our scans were run using both log and flat priors to ensure a complete coverage of the parameter space,  gathering $\sim 10^6$ samples from $\sim 10^8$ likelihood evaluations.  Samples have been thinned by a factor of 10, focusing our search to regions of the parameter space that were not clearly ruled out by LHC-Run I constraints. This produced $10^5$ representative samples to which the LHC SUSY searches have been applied. The ensuing $\sim 10^4$ samples that pass LHC run I constraints are displayed in figure~\ref{fig:results}. 

%%%%%%%%%%%%%%%%%%%%%%%%%%%%%%%%%%%%%%%%
%%%%%%%%%%%%%%%%%%%%%%%%%%%%%%%%%%%%%%%%

\bigskip
\section{Results}
\label{sec:results}

Our global fits identify two distinct viable solutions in the pMSSM parameter
space (figure~\ref{fig:results}): the first exhibits a WIMP mass of $\sim 80-100$ GeV, with the neutralino
annihilating to $WW$ with a 95\% branching ratio. The second solution has a
larger neutralino mass, $\sim 180-200$ GeV, and 87\% $\bar{t}t$ annihilation
final states. The overall best-fit point is in the $WW$ region has $-2\ln \like
\equiv \chi^2 = 122.0$. This is for a fit with 21 free parameters, and 125
Gaussian data points (we do not include limits as their $\chi^2$ is normalized
to 0 whenever the constraint is satisfied), so we adopt 104 degrees of freedom\footnote{We emphasize that the calculation of the number of actual degrees of freedom is not trivial. One would have to consider the number of {\em active} data points, as well as the number of {\em effective} parameters in the model. This can only be done properly via extensive Monte Carlo simulations of the data. The simple counting argument we adopt is meant to be representative of what one would get in the simplest scenario.}.
Our best fit thus has a $p$-value of 0.11 versus a $\chi^2 = 127.6$ and  a
$p$-value of 0.06 for the $\bar{t}t$ solution.

It is important to notice that including theoretical uncertainties in the GC
fit is crucial in achieving reasonable $p$-values. The 10\% theoretical
uncertainty advocated in~\cite{Caron:2015wda} is a reasonable reflection of the
differences in the predicted spectra between current numerical codes. However,
in absence of such an uncertainty, the quality of our global fit would degrade
to 0.023 and 0.008 for the two best-fit points, respectively.

\begin{figure*}[t!]
\centering
\includegraphics[trim=0cm 0cm 0cm 0cm, clip=true, scale=0.75]{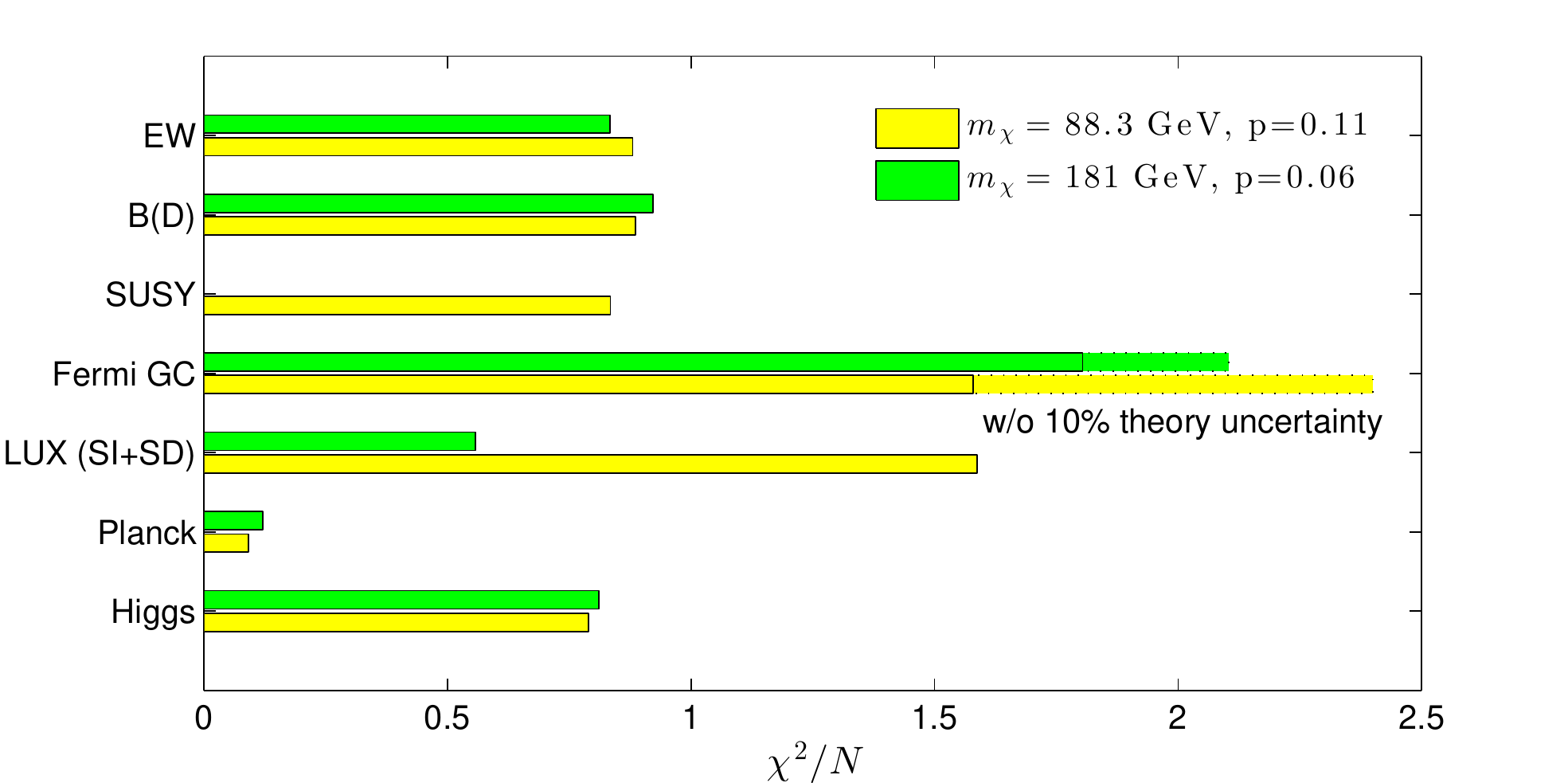}
\caption{Contribution to the overall $\chi^2$ for the two best-fit points, grouped by type of observable (see section~\ref{sec:setup} for details). The pulls have been normalized by the number of data points in each group, $N$, to facilitate a visual comparison.  The dashed bars show the Galactic Centre \Fermi\ likelihood contribution when the 10\% theoretical uncertainty is neglected, which would degrade the $p$-values to 0.023 (for the $m_\chi = 88. 3$ GeV solution) and 0.008 (for the $m_\chi = 188$ GeV model).}
\label{fig:pulls}
\end{figure*}

The contribution to the overall $\chi^2$ for the two best-fit points from different types of observables are plotted in figure~\ref{fig:pulls}. The pulls have been normalized by the number of data points in each group, $N$, to facilitate a visual comparison. We notice that the $\chi^2$ per data point is distributed fairly evenly across observables. There is a slight preponderance in the $\chi^2$ contribution coming from the GC \Fermi\ fit (with $N=24$ bins), which is exacerbated if one neglects the 10\% theoretical uncertainty in the DM spectra (dashed bars). The contribution to the pull from the LUX likelihood comes almost exclusively from the SD neutron cross-section limit, as the SI constraint is easily satisfied by our best-fit points.

Let us now analyze in more detail both type of solutions.

\bigskip

\emph{In the $WW$ region,}  model points providing a better fit have a neutralino mostly bino-like ($\sim$ 80 -- 90\%) with a similar fraction of both wino and higgsino. Besides we find points in which neutralinos can be dominantly higgsinos with a bino fraction as small as $\sim$ 10\%. Those provide a worse fit though, because a large higgsino composition, basically, implies a large 
annihilation cross section which drops the relic density below the Planck limit leading to a tension with the \Fermi~GeV excess due to the scaling Ansatz given in eq.~\eqref{eq:scaling}.

Analogously to Ref.~\cite{Caron:2015wda}, also in the present work, we find 
a WW solution with a bino-higgsino neutralino composition of about 84 -- 92 GeV.
However, in our work, this solution has a slightly worse $\chi^2$ than in Ref.~\cite{Caron:2015wda}. 
The reason for this is that we adopt slightly different values of the form-factors for the computation of the SD cross-section. 
Since such a WW solution is now just below the WW IceCube limits, it is punished a bit in the likelihood. 

Typically, since third generation squarks enter into 
the SI cross section at loop level 
through LSP-gluon effective interactions~\cite{Drees:1993bu},
their contribution can be comparable to tree-level effective interactions
 mediated by squarks of the first two families when they are light. 
In this type of solutions, because current constraints allow sbottoms of a few hundred GeV, 
their contribution can be sizable and, indeed, cancels out the Higgs exchange 
contributions. This effect allows to relax the tension with LUX data when 
Higgs exchange contributions are large. 

In terms of the impact of LHC Run-I data, we notice that first and second generation squarks as well as gluinos are decoupled. The stops and the heavier sbottom mass eigenstates are also kinematically inaccessible at the LHC Run-I energies. However, the lighter sbottom eigenstate with a mass around 400 GeV can be produced at the LHC at a considerable rate but the sbottoms evade detection from third generation searches due to complicated cascade decays. The sparticles in the electroweak sector are relatively light. There is a large mass splitting between the SU(2) doublet and singlet sleptons. The SU(2) doublet sleptons are light with masses around 250 GeV and narrowly evade detection in searches for two lepton and large missing transverse momentum final states. The lightest wino-like chargino and the second lightest neutralino escape detection since they are almost mass degenerate with the bino-like neutralino.  The production rate of the higgsino eigenstates is too small to yield an observable signal at the LHC run I.

\medskip

Results about the annihilation cross-section are shown in the top left panel of figure~\ref{fig:results}. One can see that the points with a better fit 
exhibit $\langle \sigma v \rangle \sim 10^{-26}$ cm$^3$/s, consistent with the results found in \cite{Calore:2014xka}.  We also show the \Fermi\ dwarfs limit for the $W^+W^-$ channel~\cite{Ackermann:2015zua}, but we emphasize that this limit has not been applied in the fit.  In order to compare with the constraint coming from \Fermi\ dwarfs observations, the annihilation cross-section needs to be rescaled by a factor $f_\chi^2$ (to account for the possibility of sub-dominant dark matter relic density, which translates according to our Ansatz in a correspondingly reduced local density in the dwarfs), which would suppress the signal well below current and future limits from dwarfs. 

Regarding DM direct detection, the top right panel of figure~\ref{fig:results} shows the spin-independent (SI) cross-section versus neutralino mass plane. In order to facilitate the visual comparison of our pMSSM models with existing and future limits, we have rescaled the theoretical cross-section by a factor $f_\chi$ (to account for models where the neutralino does not make up all of the cosmological relic density) and a factor $g_\chi \equiv \rho_0/0.3$ GeV/cm$^3$. This accounts for the fact that the local density we have used to predict the number of counts for LUX, $\rho_0$, is a nuisance parameter which is generally different from the value assumed by the LUX collaboration in deriving their limit~\cite{Akerib:2013tjd}, namely 0.3 GeV/cm$^3$. 

The points that appear above the nominal 95\% exclusion limit from LUX cannot be excluded because of a combination of effects: (i) our LUX likelihood is slightly less stringent than what has been published by the LUX collaboration (and depicted in figure~\ref{fig:results}), (ii) our likelihood function for LUX allows for values of the cross section above the 95\% limit to be included (albeit penalized by a smaller likelihood value), and (iii) the global likelihood function can -- to an extent -- compensate for a poor fit to LUX data by gaining an improvement from other data sets.  

Since squarks are heavy, the contribution coming from the exchange of a CP-even Higgs dominates and therefore the SI cross-section scales as $\propto |N_{11} N_{13/14}|^2$. 
Because, as noticed above, the higgsino fraction in this region is not negligible, it can be large. Indeed, there are model points above the LUX limit allowed by the scaling Ansatz  we apply to the local dark matter density and to some extend due to the fact that we vary the local dark matter density. 
Another interesting feature is that the SI cross-section spans down to $\sim 10^{-14}$ pb. That is possible because the heavy CP-even Higgs contribution can 
be sizable and cancellations with the lightest Higgs channel might occur \cite{Mandic:2000jz}. 

In the bottom panels of figure~\ref{fig:results} we display the spin dependent (SD) cross-section for 
scattering off neutrons/protons (left/right panels).  While in the case of SI interactions the contributions for proton and neutron are comparable, the SD cross-sections may differ significantly.
However, we find a tight correlation between the SD cross-sections for scattering off neutrons and those off protons in our model points. 
The SD cross-section is dominated by the exchange of a Z boson and therefore the SD cross-section 
is largely determined by the higgsino content of the neutralino and likewise for the SI cross-section it can be sizable in this region as it can be seen in both panels. 

In terms of dark matter direct detection experiments, at present,  LUX data represents the strongest constraint on the SD-neutron scattering cross-section 
because the Xenon contains neutron-odd isotopes therefore we overlay the LUX constraint properly rescaled as for the SI case. 
For the SD-proton scattering cross-section, IC-79 represents the strongest current constraint for the particular case when the neutralinos annihilate to  a W$^+$W$^-$ final state so we show the IC-79 90\% CL upper limit~\cite{Aartsen:2012kia} , and we have rescaled the value of the SD cross-section by a factor $f_\chi$. This assumes that equilibrium between capture and annihilation is reached in the Sun, which is a good assumption for the bulk of the models shown here. In fact, in the region where the neutralino annihilates mainly to $WW$ and to $ZZ$ to a lesser extend, the IC-79 limits apply here and disfavors a large number of model points.

One can see that there are model points above LUX and IC exclusion lines. In those, the higgsino component of the neutralino is dominant over the gaugino one 
leading to a large SD cross-section for both neutrons and protons. Those points still provide a reasonable fit to the data due to two effects, first because neutralinos with a large higgsino component yields to a relic density sensible below the Planck bound and therefore the scaling Ansatz applies and second  because the local dark matter density is a nuisance parameter in our analysis. Beside there is a small fraction of points above the nominal IC-79 $WW$ limit with a branching ratio to $WW$ and $ZZ$ low enough to evade the stringent IC-79 bound.

\bigskip

\emph{In the $\bar{t}t$ region,} points providing a better fit have a 
neutralino dominantly bino-like ($\sim 90\%$) with a $\sim 10\%$ of higgsino. Those points have the
characteristic that the neutralinos annihilate to top quark pairs via an exchange of a right-handed stop which is 
relatively heavy ($\sim$ 1 TeV). This is possible because on top of the non-helicity suppression the neutralino-stop-top 
coupling component, which is proportional to the top quark Yukawa coupling, is sizeable due 
to the  non-vanishing higgssino fraction of the neutralino.  
We also found points in which the right-handed stops are light ($\sim$ 300 GeV) being almost bino-like 
(also found in \cite{Caron:2015wda}). However those provide a worse fit to the $\brbsgamma$ data because higgssino-stop loops have a sizable positive contribution which leads to values above 
the experimental constraint.

Annihilation into $t \bar{t}$ 
through a t-channel stop exchange requires, 
for bino-like neutralinos, stops with masses 
of a few hundred GeV. On the other hand, relatively light stops are not 
able to lift the tree-level Higgs mass to fit a 125 GeV Higgs. 
In order to enhance the DM annihilation cross sections to match the 
Planck measurement, the neutralino coannihilates with sneutrinos which 
have to be about the same mass as the DM particle ($\sim$ 200 GeV). This 
induces the splitting in the left/right sleptons spectrum.

These benchmark points are characterized by left handed slepton next-to-lightest supersymmetric particle with masses above 200 GeV. The higgsinos have masses around 260 GeV. Since the resulting mass differences between the bino-like LSP and the sleptons/higgsinos are small, the final states are rather soft and thus the detection is suppressed in events with dilepton or trilepton final states and large missing transverse energy momentum. The production rate of the sbottom is quite suppressed and hence avoids detection. The spectrum of the remaining supersymmetric particles is decoupled.

\medskip

The phenomenology of the model points in the $\bar{t}t$ region in terms of DM detection is similar to the $WW$ one. The main 
difference is in that the IC-79 limits does not apply and therefore larger SD
cross-sections are possible.
The main difference is that the higgsino composition is not as large as in the WW type of solutions and therefore the SD cross sections for neutron interactions are below the LUX current sensitivity. 

\bigskip

\begin{figure*}[t!]
\includegraphics[width=0.47\linewidth]{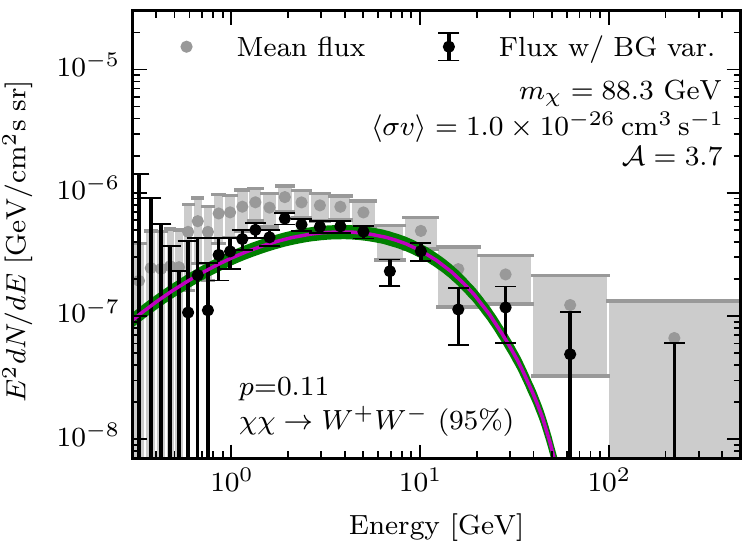} %WW
\includegraphics[width=0.47\linewidth]{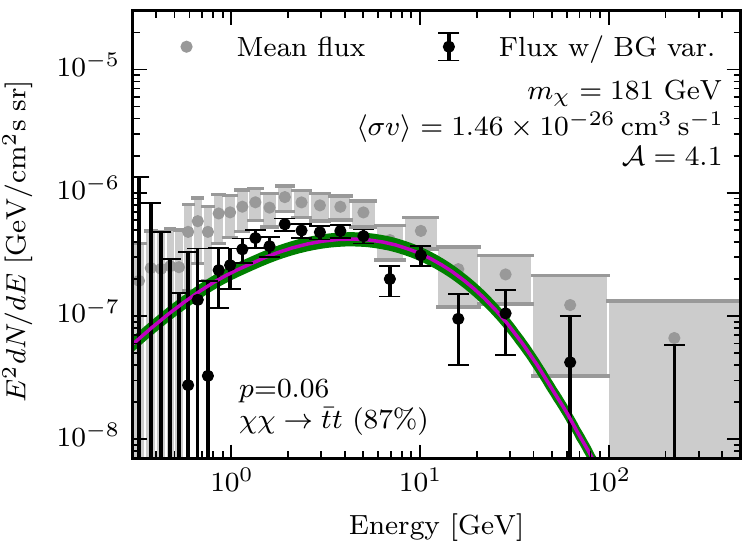} %tt
\caption{Spectral energy distribution of the \Fermi-GeV excess data.  \emph{Grey dots} show \Fermi~GeV excess mean values w.r.to background model variations, together with associated systematic uncertainties (\emph{grey boxes}). \emph{Black dots} represent the excess for a variation of the Galactic
diffuse emission contribution (within its systematic uncertainty).
The solid lines show the prediction for the pMSSM
models that give the best fits in the two regions of figure~\ref{fig:results} and have dominant 
annihilation channel into W bosons or top quarks (\emph{magenta lines} in the left 
and right panel, respectively).  The \emph{green lines} indicate the adopted $10\%$
systematic uncertainties in the spectra.  Furthermore, $\mathcal{A}$
denotes the required boost-factor with respect to a generalized Navarro-Frenk-White DM density 
profile with inner slope $\gamma$=1.26. The $p$
denotes the $p$-value of the global fit (including all data).}
\label{fig:spectra} 
\end{figure*}

Lastly, in figure~\ref{fig:spectra}, we show the spectrum of the \Fermi~GeV excess together with 
the systematic uncertainties associated with the galactic diffuse emission modeling~\cite{Calore:2014xka}.
We compare the data with the spectra of the pMSSM model points 
giving the best global $p$-value in the two regions identified in Figure~\ref{fig:results}. 

It is apparent from Figure~\ref{fig:spectra} that the
best-fit DM spectra are systematically offset from the mean values of the
Galactic center excess spectrum (gray dots and boxes), by about 1 to 2 sigma, and do not provide a
good fit to the data at first sight.  However, since the systematic
astrophysical uncertainties, indicated by the gray boxes ($\pm1\sigma$) are
\emph{correlated}, this still provides an acceptable fit to the data.  To
\emph{illustrate} this point, we show with black dots and error bars the excess
spectrum where we moved all data points systematically down, according to the
freedom allowed by the covariance (the error bars show now statistical errors
only).  Together with the 10\% uncorrelated systematic modeling uncertainty
that we adopted for the DM signal, this provides a reasonable fit to the data,
with $p$-values, whereas without the DM signal modeling uncertainties, the
$p$-values would be prohibitively small (see Figure~\ref{fig:pulls} above).

%%%%%%%%%%%%%%%%%%%%%%%%%%%%%%%%%%%%%%%%
%%%%%%%%%%%%%%%%%%%%%%%%%%%%%%%%%%%%%%%%
\section{Discovery potential: dark matter detection experiments and LHC run II}
\subsection{Implications for direct and indirect dark matter searches}
\label{sec:implicationsDDID}
% direct detection
Generally, spin-dependent and spin-independent scattering cross sections are driven by
the higgsino content of the neutralino. 
Therefore, as explained above, the sizable higgsino fraction of points in the two best-fit regions imply large SI cross-sections
that makes the direct detection prospects promising, although the SI cross-section range  spans down to 
$\sim 10^{-14}$ pb due to cancellations with the lightest Higgs.
In figure~\ref{fig:results}, top right panel, we also display the projected sensitivity limit (defined as the 90\% CL exclusion limit) for the Xenon1T experiment and an
 hypothetical liquid Xe detector with 500 t$\times$ yr exposure~\cite{2015arXiv150608309S}. The latter experiment essentially
saturates the ultimate detection floor set by coherent neutrino scattering~\cite{Billard:2014yka}. 
Xenon1T data will be crucial in discovering or firmly ruling out models belonging to the $\bar{t}t$ island and will
probe a significant fraction of the parameter space preferred by the first type of solutions. 
In figure~\ref{fig:results}, bottom left panel, we overlay the projected 90\% exclusion limit for Xenon1T~\cite{2015arXiv150608309S}.
We find that Xenon1T will be able to prove the entirety of the SD neutron scattering cross-section parameter space favoured by our models. \footnote{
We note that the Xenon1T exclusion limits in~\cite{2015arXiv150608309S} are obtained by applying a scaling factor  
derived from the comparison between SI and SD results of Xenon100 and result to be slightly stronger than the limits quoted in
ref.~\cite{Garny:2012it}, which instead adopts a 60 times improvement w.r.to XENON100 (still Xenon1T will be able to entirely probe our identified parameter space).
}
A multi-ton experiment with 500 t$\times$ yr exposure would reach sensitivities a couple of 
order of magnitudes smaller than the smallest SD neutron cross-sections found in our scan. 

The bottom right panel of figure~\ref{fig:results} shows that our best-fit  points easily evade the constraint set by IC-79 on the SD proton cross-section in the implementation we adopted in this paper. However, an event-level implementation of the likelihood (including the events' energies~\cite{Scott:2012mq}) would increase the constraining power of the IC-79 limit, to the point that some of the surviving models could be probed~\cite{IC79future}. 

Finally, as for indirect detection, the preferred parameter space is mostly
out of reach even for the future 10-yr \Fermi\ analysis of dwarf spheroidal
 galaxies~\cite{Bringmann:2012ez}, where an further improvement of current sensitivities by a factor
 2 -- 3 can be expected.  Note that points with a large annihilation cross
section as shown in the top right panel of figure~\ref{fig:results} usually correspond to
suppressed relic densities, making these points hard to detect due to the $f_\chi^2$ factor suppression of their signal.

%%%%%%%%%%%%%%%%%%%%%%%%%%%%%%%%%%%%%%%%
%%%%%%%%%%%%%%%%%%%%%%%%%%%%%%%%%%%%%%%%
\subsection{Prospects for detection at the LHC run II}
\label{sec:LHCII}

\begin{figure*}[t!]
\centering
\includegraphics[trim=4cm 3.7cm 5.5cm 6cm, clip=true, scale=0.6]{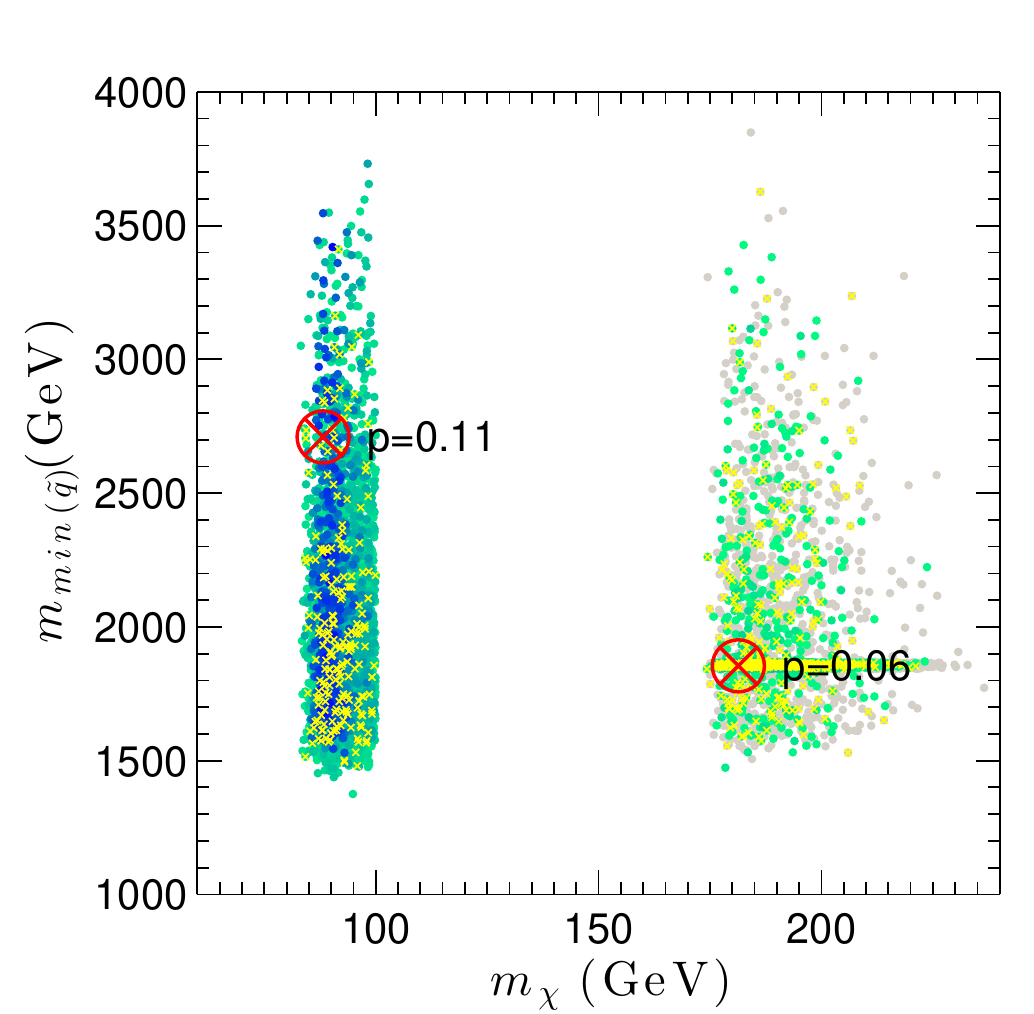}
\includegraphics[trim=3.6cm 3.7cm 3cm 6cm, clip=true, scale=0.6]{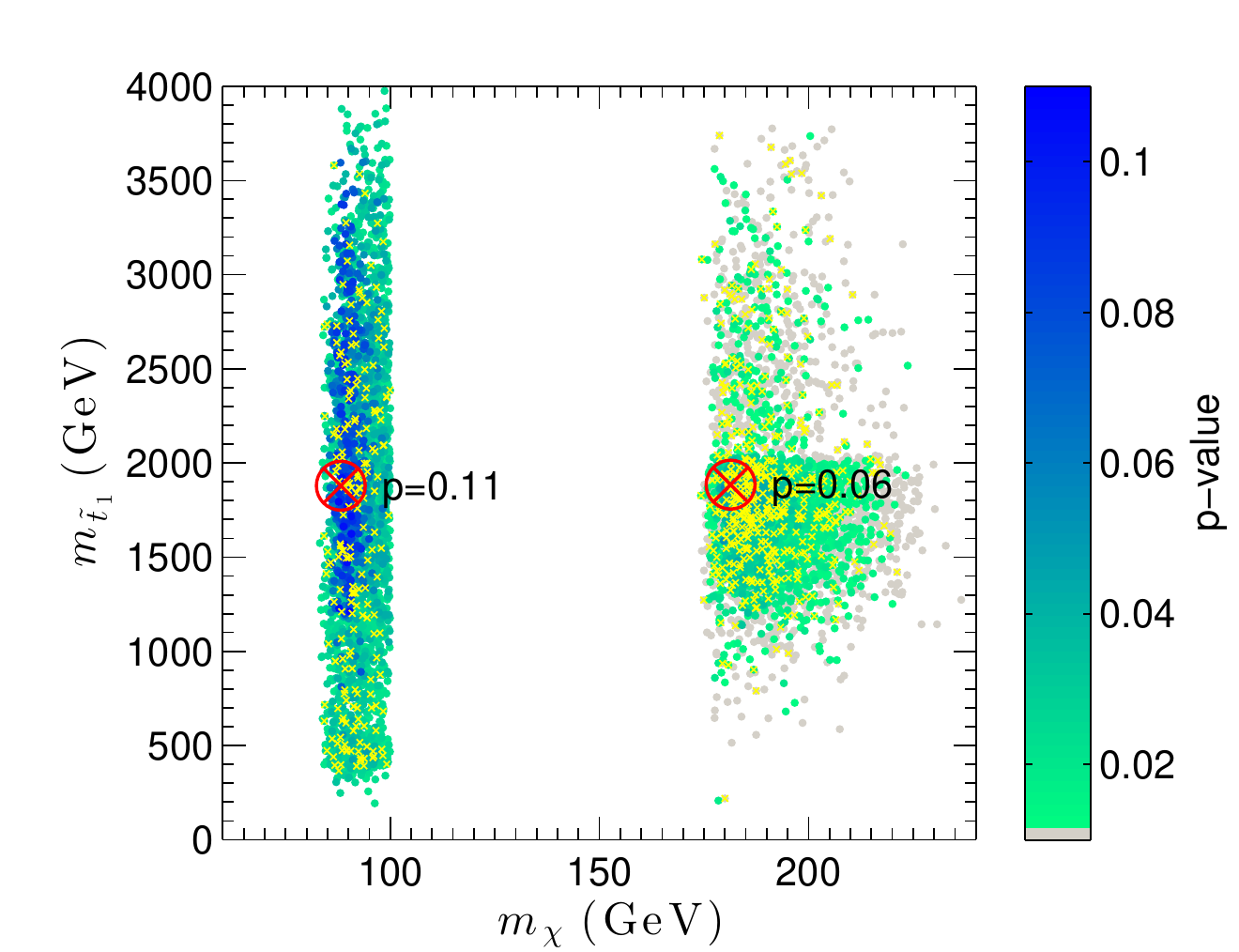}  \\ 
\includegraphics[trim=4cm 3.7cm 5.5cm 6cm, clip=true, scale=0.6]{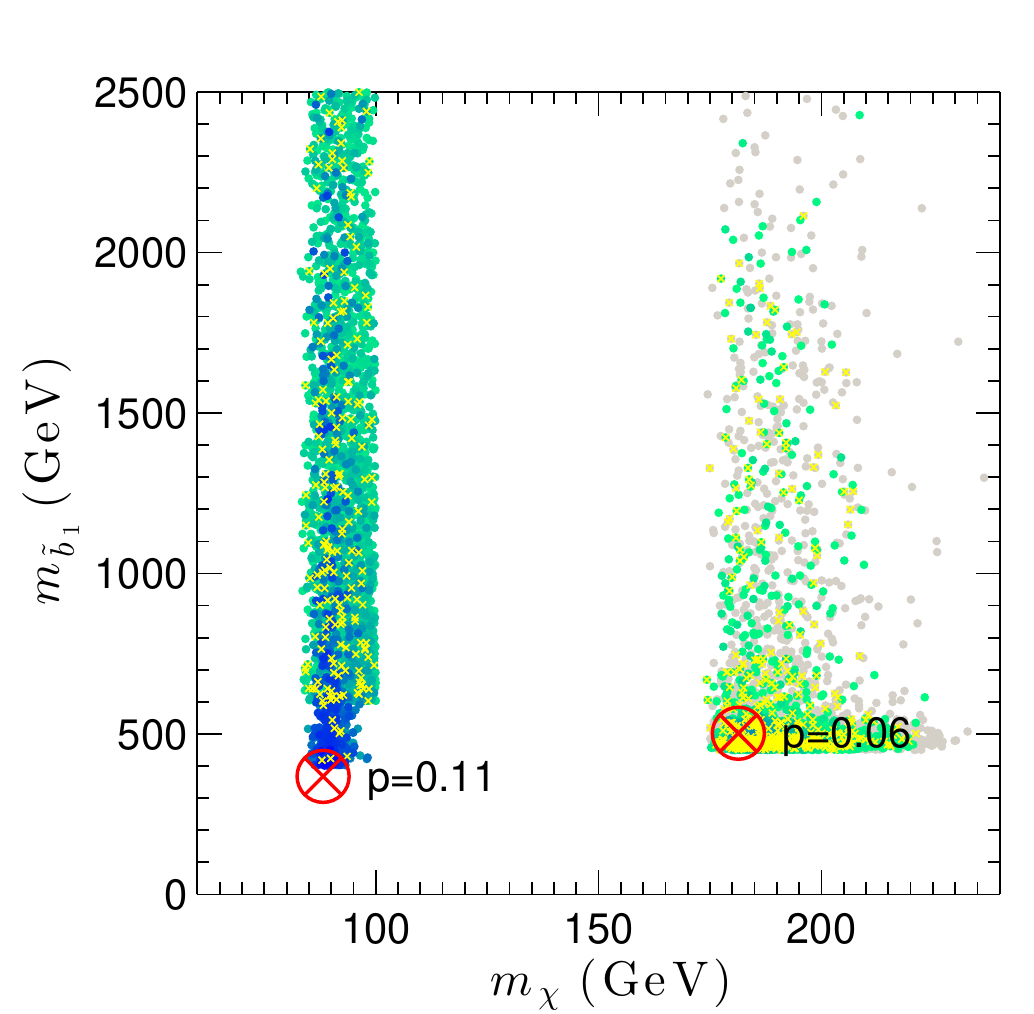}  % left, bottom, right and top \\
\includegraphics[trim=3.6cm 3.7cm 3cm 6cm, clip=true, scale=0.6]{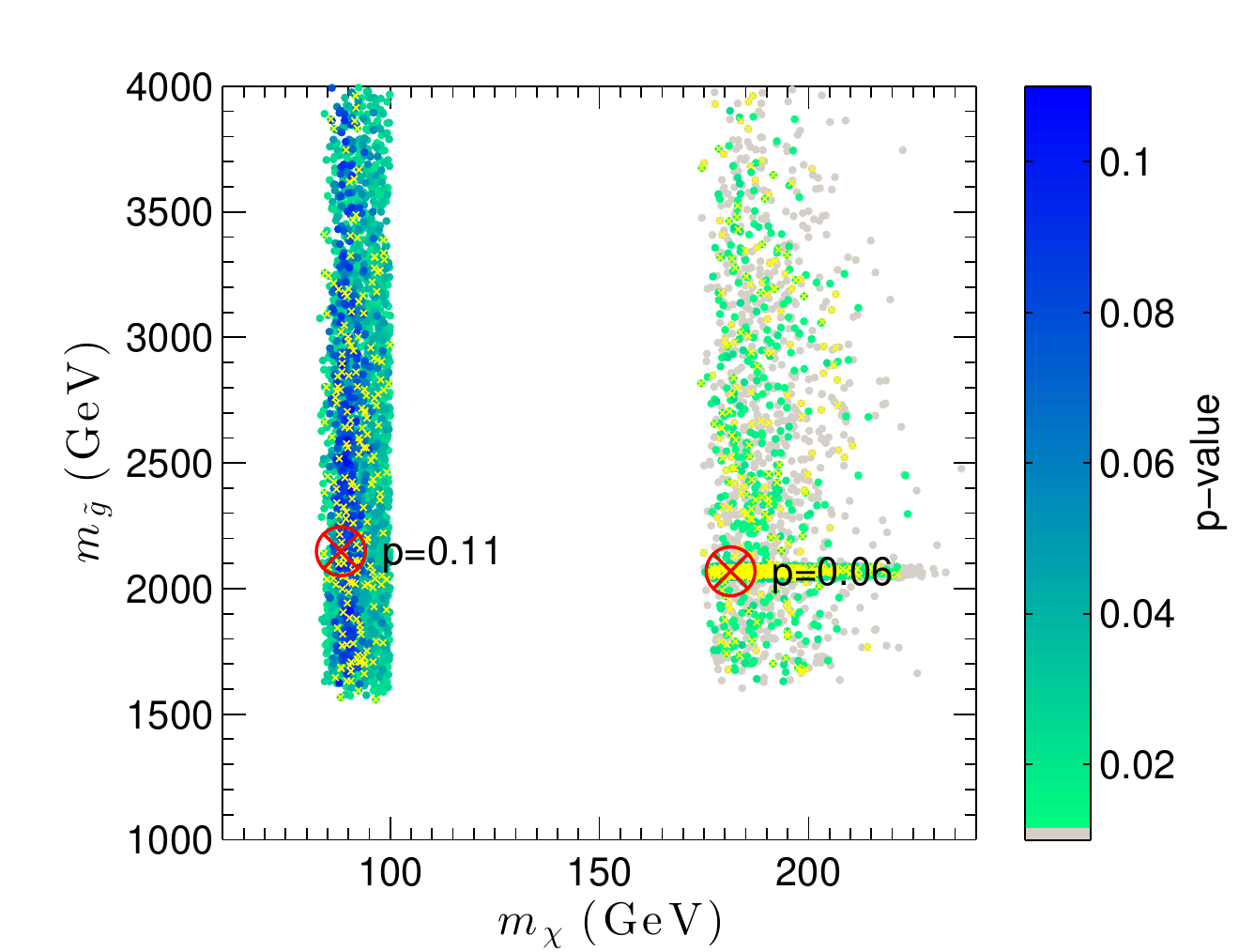} \\
\includegraphics[trim=4cm 3.7cm 5.5cm 6cm, clip=true, scale=0.6]{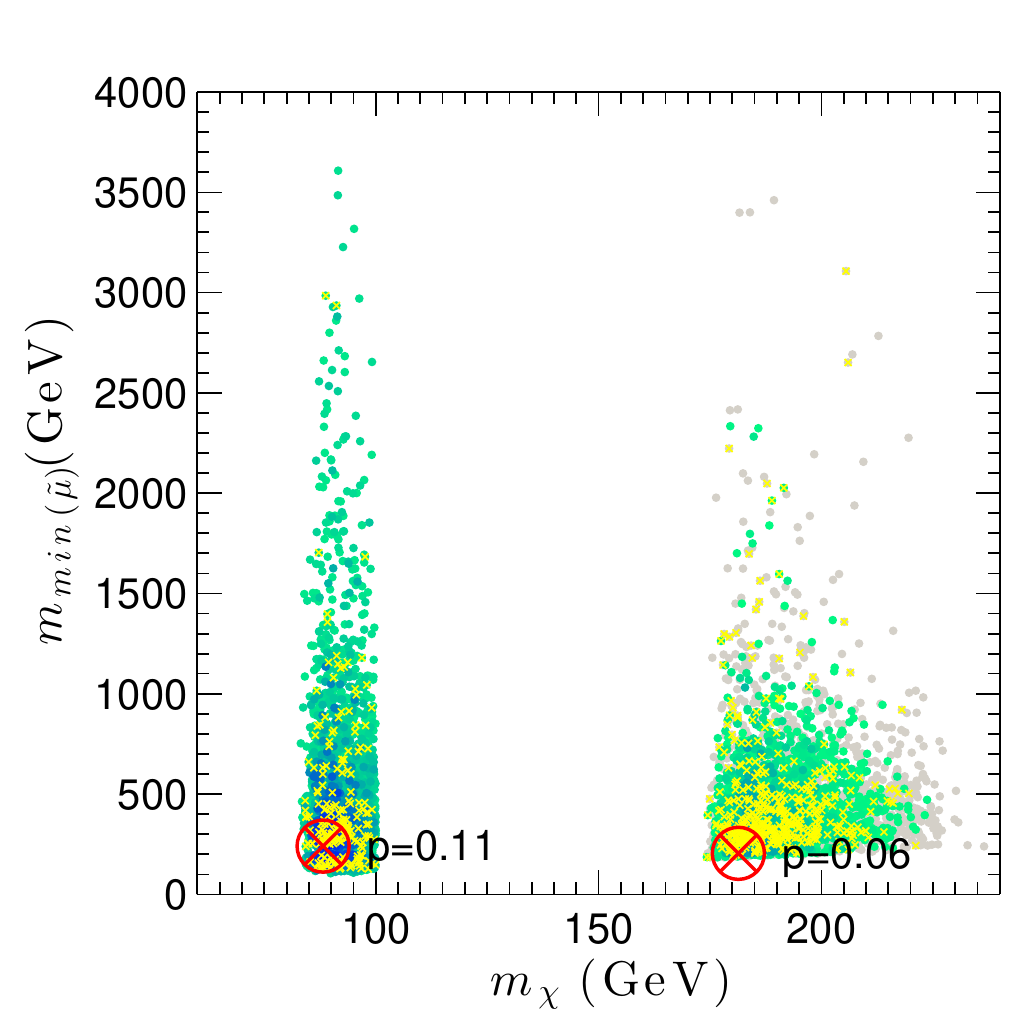} 
\includegraphics[trim=3.6cm 3.7cm 3cm 6cm, clip=true, scale=0.6]{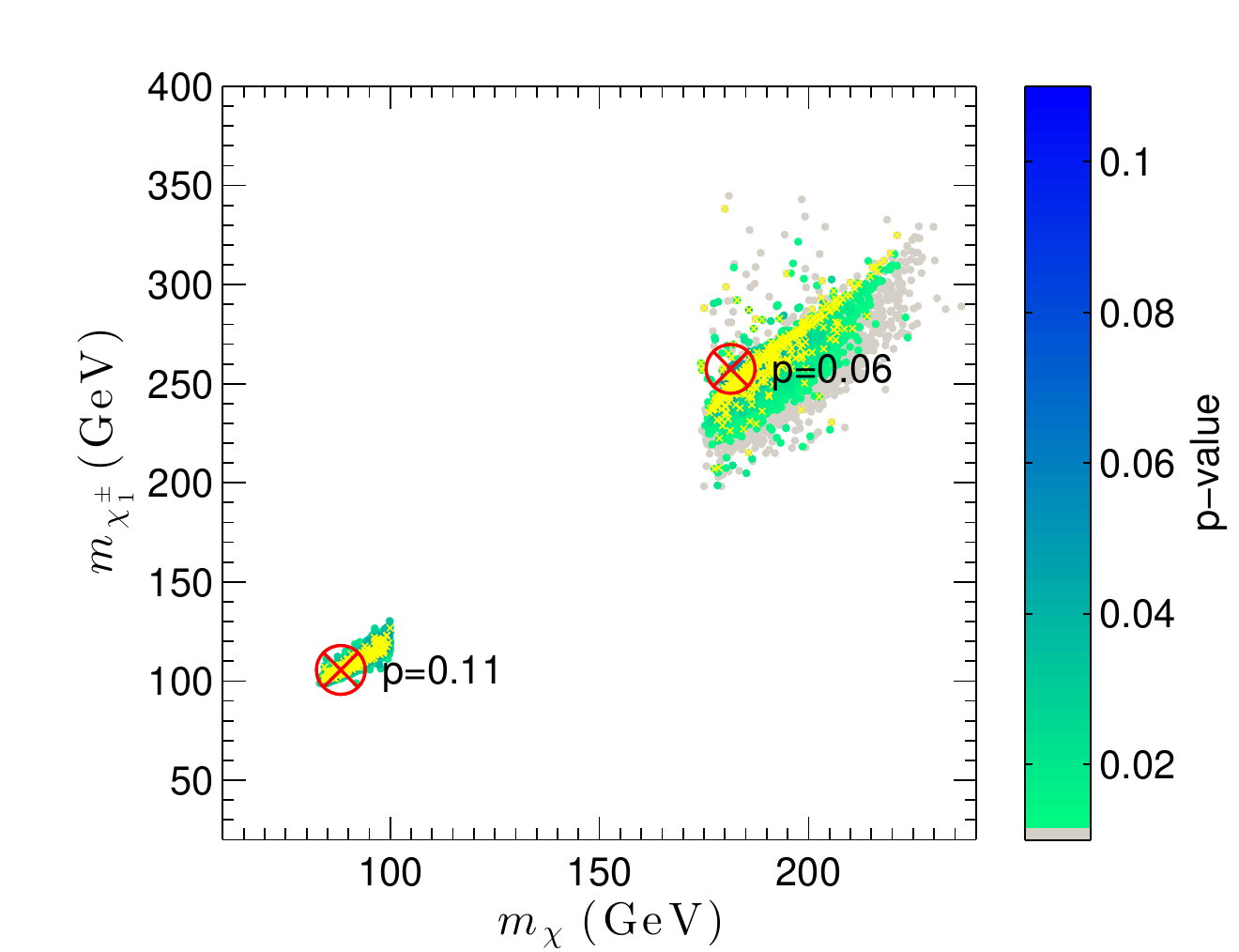} \\% left, bottom, right and top \\

\caption{2D map of the $p$-values of our fit, showing the lightest squark mass of the first and second generation (top left), the lightest stop (top right), the lightest sbottom (middle left), the gluino (middle right),
 the lightest smuon (bottom left) and the lightest chargino (bottom right)
vs the neutralino mass. The yellow overlay shows points within $2\sigma$ of the {\em Planck} relic density value. 
}
\label{fig:resultsLHC}
\end{figure*}

The models yielding the best $p$-values show
several interesting properties relevant for LHC searches.
In the following we briefly discuss
the discovery prospects depending on the
produced SUSY particles. Sparticle mass distributions and MSSM
parameters relevant for LHC searches are shown in figure~\ref{fig:resultsLHC}.

\paragraph{{\it Light squarks}:} In the top left panel of Fig.~\ref{fig:resultsLHC} we plot the lightest first/second generation squark mass vs the neutralino mass. 
The first and second generation squarks have masses $>1400$~GeV, i.e. above 
the usual run I constraints. About $70\%$ of the
models have squark masses below $2000$~GeV
\footnote{The expected reach of the ATLAS and CMS experiments
for full MSSM models is not estimated. We compare with mass scales
which are likely excluded with HL-LHC data for many MSSM scenarios.}.
The upcoming run II
searches for light squarks will exclude many scenarios.
In almost all models the left- and right-handed squarks have different branching
ratios and tend to decay to the heavy neutralino and chargino states.
Cascade decays including W, Z and Higgs bosons are common.

\paragraph{ {\it Stop}:} In the top right panel of Fig.~\ref{fig:resultsLHC} we plot the stop mass vs the neutralino mass. 
Some models have light stops with masses down to 200 -- 300 GeV decaying
to chargino and a b-jet. The neutralino has a mass of around 95 GeV.
These models are not excluded by current LHC searches~\cite{Aad:2014kra}.
Another interesting region also found in ~\cite{Caron:2015wda}
has a stop mass of around 200 -- 220 GeV and a mass of the lightest neutralino
around 180 GeV. A slight excess in the ATLAS data prevents exclusion
with run-1 data in this region ~\cite{Aad:2014kra}.
These solutions will likely be tested with early run II analyses.
Other solutions yield much heavier stop masses decaying
predominantly into the heavier neutralino and chargino states. Dedicated
searches for such decays are important.
%#example slhafile_shit_1489.slha

\paragraph{{\it Sbottom}:} In the central left panel of Fig.~\ref{fig:resultsLHC} we plot the sbottom mass vs the neutralino mass.  Several model points have a sbottom mass as low
as 400 GeV. The points are not excluded in our procedure
due to multi-step cascade decays involving heavy neutralinos.

Typically, the lighter sbottom state typically has masses around 
400 GeV while the lightest neutralino is mostly bino-like with a mass around 90 GeV. The second 
lightest wino-like neutralino lies around 107 GeV. The two heavier neutralinos are higgsino 
dominated states and have masses around 250 GeV.

If the sbottoms predominantly decayed into a bottom quark and the lightest neutralino, these
 benchmark points would clearly be excluded. However, the bottom squark decays into all four 
 neutralino as well as the lighter chargino mass eigenstates with comparable rates. The decay 
 modes of the neutralino and chargino mass eigenstates are relatively complex and hence the
  limits from simplified sbottom searches do not apply. For instance, the second lightest neutralino 
  eigenstate has large hadronic three body decay modes into the lightest neutralino via off shell Z bosons. 
  Moreover, the second lightest neutralino can radiatively decay into a photon and the lightest neutralino. 
  The corresponding lighter chargino eigentstate has relatively large leptonic three body decay modes via off shell W bosons. 
  Finally, the two heaviest higgsino dominated neutralino mass eigenstates mainly decay into electroweak 
  gauge bosons and the lighter electroweakino states. As a consequence, many events have leptons in the 
  final state which are vetoed in the searches for direct sbottom production. In addition, the higher final state
   multiplicity tends to soften the net missing transverse momentum distribution compared to scenarios with
    direct sbottom decays into a bottom quark and the LSP. As a result fewer events pass the selection cuts 
 of the relevant sbottom searches. We explicitly tested those light sbottom scenarios against experimental 
 searches at the LHC with the computer tool CheckMATE and confirmed that those points were allowed.

As seen also for the light squarks about 70\% of the models would
be excluded with a sbottom limit of $\approx 1000$~GeV.

\paragraph{ {\it Gluinos}:} In the central right  panel of Fig.~\ref{fig:resultsLHC} we plot the sbottom mass vs the neutralino mass.  Gluinos have masses $>1600$~GeV.
An upcoming early run II exclusion
on gluinos with a mass up to 2500 GeV would exclude about $15\%$ of 
these models. Run II searches for first and second generation 
squarks will thus likely be more constraining.

\paragraph{{\it Sleptons}:} In the bottom left panel of Fig.~\ref{fig:resultsLHC} we plot the lightest smuon mass vs the neutralino mass.  The lightest smuons found in the best fit models
have masses $<400$~GeV in about $60\%$ of the best fit points.
This makes searches for smuons in run II very sensitive to 
these solutions.

\paragraph{ {\it Chargino/neutralino}:} In the bottom right panel of Fig.~\ref{fig:resultsLHC} we plot the chargino mass vs the neutralino mass. Due to the GeV excess likelihood 
several neutralinos
and charginos are typically light, the lightest having a mass fixed to
$80-100$~GeV for the $WW$ solutions and $180 - 200$~GeV for the $\overline{t}t$ solutions. The higgsino component
in the $\overline{t}t$ and $WW$ solutions typically involves that $\mu$
is only slightly larger than these mass scales, leading to 2 more
neutralinos and the light chargino
at masses around 100 GeV ($WW$) or 200 -- 300 GeV ($\overline{t}t$).
These states are often mass compressed with the lightest neutralino
which makes the solutions evade LHC chargino/neutralino searches so far.
Dedicated chargino/neutralino searches will have sensitivity
to most models, e.g. by a mono-jet and soft lepton search as proposed
also in ref.~\cite{Caron:2015wda}.
The Wino mass scale is quite unconstrained and lies between 100 GeV
and 1.5 TeV. The Wino will decay to lighter states yielding final states with
Z, W or Higgs bosons.

\paragraph{ {\it Heavy Higgs}:}
About $50\%$ of the best models have $m_A<800$~GeV making searches
for heavy Higgs bosons very sensitive.
Several chargino and neutralino states are light and have
a large coupling to $A/H/H^{\pm}$. Consequently heavy Higgs decays
to charginos and neutralinos
can have huge branching ratios up to $30\%$ competing with top and
bottom decays. Dedicated searches for heavy Higgs bosons decaying 
into final states with $W/Z/h$ with missing transverse momentum
would help to constrain these scenarios.

%%%%%%%%%%%%%%%%%%%%%%%%%%%%%%%%%%%%%%%%
%%%%%%%%%%%%%%%%%%%%%%%%%%%%%%%%%%%%%%%%
\section{Conclusions}
\label{sec:conclusions}
In this paper we addressed the issue of finding model points in the pMSSM
that might explain simultaneously the large set of independent data we gathered from astrophysics,
cosmology and high-energy particle physics.
We showed  that no tension exists between currently available particle physics constraints and
the interpretation of the \Fermi~GeV excess in terms of dark matter annihilation in the framework of 
the pMSSM. Furthermore, we found evidence for two regions 
that are able to explain the gamma-ray data, while being consistent with other various experimental
constraints: (a) a first region where the neutralino is mostly bino-like and the dominant annihilation
channel today is 95\% into $W$ bosons pairs
and (b) a second region where the the annihilation into top-quark pairs dominates and the neutralino is again mainly bino-like.
We showed that these models are very appealing since they will be soon in the reach of the next generation of direct 
detection experiments -- Xenon1T will probe the entirety of the best-fit regions thanks to its sensitivity to spin-dependent neutron cross-section -- and of the LHC run II, in particular through searches for 
charginos and neutralinos, squarks and light smuons.

%%%%%%%%%%%%%%%%%%%%%%%%%%%%%%%%%%%%%%%%
%%%%%%%%%%%%%%%%%%%%%%%%%%%%%%%%%%%%%%%%
\bigskip
{\bf Acknowledgments.} 
We warmly thank T.~Bringmann,  Marco Selvi, Christopher McCabe, Pat Scott and Tim Stefaniak for useful discussions, Marc Schumann and Marco Selvi for providing us with direct detection limits and William Parker for assistance with some figures.  
G.B. (P.I.) and F.C.~acknowledge support from the European Research Council
through the ERC starting grant WIMPs Kairos.
R. RdA, is supported by the Ram\'on y Cajal program of the Spanish MICINN and also thanks the support of the Spanish 
MICINN's Consolider-Ingenio 2010 Programme 
under the grant MULTIDARK CSD2209-00064, the Invisibles European ITN project (FP7-PEOPLE-2011-ITN, 
PITN-GA-2011-289442-INVISIBLES and the 
``SOM Sabor y origen de la Materia" (FPA2011-29678) and the ``Fenomenologia y Cosmologia de la Fisica mas 
alla del Modelo Estandar e lmplicaciones Experimentales 
en la era del LHC" (FPA2010-17747) MEC projects.
The work of J. S. Kim has been partially supported by the MINECO, Spain, under contract FPA2013-44773-P;
 Consolider-Ingenio CPAN CSD2007-00042 and the Spanish MINECO Centro de excelencia Severo Ochoa 
 Program under grant SEV-2012-0249. R.T. acknowledges partial support from an EPSRC ``Pathways to Impact'' grant. 
C.W.~was supported by the Netherlands Organization for Scientific Research
(NWO) through a Vidi grant.
We gratefully acknowledge the use of the Cartesius supercomputer (Amsterdam) and of Imperial College London's HPC. This work was supported by Grant ST/N000838/1 from the Science and Technology Facilities Council. 

\bibliographystyle{JHEP}
\bibliography{pMSSMGC.bib}

\providecommand{\href}[2]{#2}\begingroup\raggedright\begin{thebibliography}{10}

\bibitem{Jungman96}
G.~Jungman, M.~Kamionkowski, and K.~Griest, {\it {Supersymmetric dark matter}},
   {\em Phys. Rept.} {\bf 267} (1996) 195--373,
  [\href{http://arxiv.org/abs/hep-ph/9506380}{{\tt hep-ph/9506380}}].

\bibitem{Bergstrom00}
L.~Bergstrom, {\it {Nonbaryonic dark matter: Observational evidence and
  detection methods}},  {\em Rept. Prog. Phys.} {\bf 63} (2000) 793,
  [\href{http://arxiv.org/abs/hep-ph/0002126}{{\tt hep-ph/0002126}}].

\bibitem{Bertone05}
G.~Bertone, D.~Hooper, and J.~Silk, {\it {Particle dark matter: Evidence,
  candidates and constraints}},  {\em Phys. Rept.} {\bf 405} (2005) 279--390,
  [\href{http://arxiv.org/abs/hep-ph/0404175}{{\tt hep-ph/0404175}}].

\bibitem{BertoneBook}
G.~{Bertone}, ed., {\em {Particle Dark Matter: Observations, Models and
  Searches}}.
\newblock Cambridge University Press, 2010.

\bibitem{Planck:2006uk}
{The Planck Collaboration}, {\it {The Scientific Programme of Planck}},  {\em
  ArXiv Astrophysics e-prints} (Apr., 2006)
  [\href{http://arxiv.org/abs/astro-ph/0}{{\tt astro-ph/0}}].

\bibitem{Goodenough:2009gk}
L.~Goodenough and D.~Hooper, {\it {Possible Evidence For Dark Matter
  Annihilation In The Inner Milky Way From The Fermi Gamma Ray Space
  Telescope}},  \href{http://arxiv.org/abs/0910.2998}{{\tt arXiv:0910.2998}}.

\bibitem{Vitale:2009hr}
{\bf Fermi/LAT Collaboration} Collaboration, V.~Vitale and A.~Morselli, {\it
  {Indirect Search for Dark Matter from the center of the Milky Way with the
  Fermi-Large Area Telescope}},  \href{http://arxiv.org/abs/0912.3828}{{\tt
  arXiv:0912.3828}}.

\bibitem{Hooper:2010mq}
D.~Hooper and L.~Goodenough, {\it {Dark Matter Annihilation in The Galactic
  Center As Seen by the Fermi Gamma Ray Space Telescope}},  {\em Phys.Lett.}
  {\bf B697} (2011) 412--428, [\href{http://arxiv.org/abs/1010.2752}{{\tt
  arXiv:1010.2752}}].

\bibitem{Hooper:2011ti}
D.~Hooper and T.~Linden, {\it {On The Origin Of The Gamma Rays From The
  Galactic Center}},  {\em Phys.Rev.} {\bf D84} (2011) 123005,
  [\href{http://arxiv.org/abs/1110.0006}{{\tt arXiv:1110.0006}}].

\bibitem{Abazajian:2012pn}
K.~N. Abazajian and M.~Kaplinghat, {\it {Detection of a Gamma-Ray Source in the
  Galactic Center Consistent with Extended Emission from Dark Matter
  Annihilation and Concentrated Astrophysical Emission}},  {\em Phys.Rev.} {\bf
  D86} (2012) 083511, [\href{http://arxiv.org/abs/1207.6047}{{\tt
  arXiv:1207.6047}}].

\bibitem{Gordon:2013vta}
C.~Gordon and O.~Macias, {\it {Dark Matter and Pulsar Model Constraints from
  Galactic Center Fermi-LAT Gamma Ray Observations}},  {\em Phys.Rev.} {\bf
  D88} (2013) 083521, [\href{http://arxiv.org/abs/1306.5725}{{\tt
  arXiv:1306.5725}}].

\bibitem{Hooper:2013rwa}
D.~Hooper and T.~R. Slatyer, {\it {Two Emission Mechanisms in the Fermi
  Bubbles: A Possible Signal of Annihilating Dark Matter}},  {\em Phys.Dark
  Univ.} {\bf 2} (2013) 118--138, [\href{http://arxiv.org/abs/1302.6589}{{\tt
  arXiv:1302.6589}}].

\bibitem{Abazajian:2014fta}
K.~N. Abazajian, N.~Canac, S.~Horiuchi, and M.~Kaplinghat, {\it {Astrophysical
  and Dark Matter Interpretations of Extended Gamma Ray Emission from the
  Galactic Center}},  \href{http://arxiv.org/abs/1402.4090}{{\tt
  arXiv:1402.4090}}.

\bibitem{Daylan:2014rsa}
T.~Daylan, D.~P. Finkbeiner, D.~Hooper, T.~Linden, S.~K.~N. Portillo, et~al.,
  {\it {The Characterization of the Gamma-Ray Signal from the Central Milky
  Way: A Compelling Case for Annihilating Dark Matter}},
  \href{http://arxiv.org/abs/1402.6703}{{\tt arXiv:1402.6703}}.

\bibitem{Calore:2014xka}
F.~Calore, I.~Cholis, and C.~Weniger, {\it {Background model systematics for
  the Fermi GeV excess}},  \href{http://arxiv.org/abs/1409.0042}{{\tt
  arXiv:1409.0042}}.

\bibitem{fermigc}
{\bf Fermi-LAT} Collaboration, S.~Murgia {\em {Talk given at the 2014 Fermi
  Symposium, Nagoya, Japan, October 20-24}} (2014).

\bibitem{Calore:2014nla}
F.~Calore, I.~Cholis, C.~McCabe, and C.~Weniger, {\it {A Tale of Tails: Dark
  Matter Interpretations of the Fermi GeV Excess in Light of Background Model
  Systematics}},  \href{http://arxiv.org/abs/1411.4647}{{\tt arXiv:1411.4647}}.

\bibitem{Abazajian:2010zy}
K.~N. Abazajian, {\it {The Consistency of Fermi-LAT Observations of the
  Galactic Center with a Millisecond Pulsar Population in the Central Stellar
  Cluster}},  {\em JCAP} {\bf 1103} (2011) 010,
  [\href{http://arxiv.org/abs/1011.4275}{{\tt arXiv:1011.4275}}].

\bibitem{Calore:2014oga}
F.~Calore, M.~Di~Mauro, F.~Donato, and F.~Donato, {\it {Diffuse gamma-ray
  emission from galactic pulsars}},  {\em Astrophys.J.} {\bf 796} (2014) 1,
  [\href{http://arxiv.org/abs/1406.2706}{{\tt arXiv:1406.2706}}].

\bibitem{Yuan:2014rca}
Q.~Yuan and B.~Zhang, {\it {Millisecond pulsar interpretation of the Galactic
  center gamma-ray excess}},  \href{http://arxiv.org/abs/1404.2318}{{\tt
  arXiv:1404.2318}}.

\bibitem{Petrovic:2014xra}
J.~Petrovic, P.~D. Serpico, and G.~Zaharijas, {\it {Millisecond pulsars and the
  Galactic Center gamma-ray excess: the importance of luminosity function and
  secondary emission}},  \href{http://arxiv.org/abs/1411.2980}{{\tt
  arXiv:1411.2980}}.

\bibitem{Carlson:2014cwa}
E.~Carlson and S.~Profumo, {\it {Cosmic Ray Protons in the Inner Galaxy and the
  Galactic Center Gamma-Ray Excess}},  {\em Phys.Rev.} {\bf D90} (2014) 023015,
  [\href{http://arxiv.org/abs/1405.7685}{{\tt arXiv:1405.7685}}].

\bibitem{Petrovic:2014uda}
J.~Petrovic, P.~D. Serpico, and G.~Zaharijas, {\it {Galactic Center gamma-ray
  "excess" from an active past of the Galactic Centre?}},  {\em JCAP} {\bf
  1410} (2014), no.~10 052, [\href{http://arxiv.org/abs/1405.7928}{{\tt
  arXiv:1405.7928}}].

\bibitem{Cholis:2015dea}
I.~Cholis, C.~Evoli, F.~Calore, T.~Linden, C.~Weniger, et~al., {\it {The
  Galactic Center GeV Excess from a Series of Leptonic Cosmic-Ray Outbursts}},
  \href{http://arxiv.org/abs/1506.05119}{{\tt arXiv:1506.05119}}.

\bibitem{Gaggero:2015nsa}
D.~Gaggero, M.~Taoso, A.~Urbano, M.~Valli, and P.~Ullio, {\it {Towards a
  realistic astrophysical interpretation of the Galactic center excess}},
  \href{http://arxiv.org/abs/1507.06129}{{\tt arXiv:1507.06129}}.

\bibitem{Bartels:2015aea}
R.~Bartels, S.~Krishnamurthy, and C.~Weniger, {\it {Strong support for the
  millisecond pulsar origin of the Galactic center GeV excess}},
  \href{http://arxiv.org/abs/1506.05104}{{\tt arXiv:1506.05104}}.

\bibitem{Lee:2015fea}
S.~K. Lee, M.~Lisanti, B.~R. Safdi, T.~R. Slatyer, and W.~Xue, {\it {Evidence
  for Unresolved Gamma-Ray Point Sources in the Inner Galaxy}},
  \href{http://arxiv.org/abs/1506.05124}{{\tt arXiv:1506.05124}}.

\bibitem{Brandt:2015ula}
T.~D. Brandt and B.~Kocsis, {\it {Disrupted Globular Clusters Can Explain the
  Galactic Center Gamma Ray Excess}},
  \href{http://arxiv.org/abs/1507.05616}{{\tt arXiv:1507.05616}}.

\bibitem{Djouadi:1998di}
{\bf MSSM Working Group} Collaboration, A.~Djouadi et~al., {\it {The Minimal
  supersymmetric standard model: Group summary report}},
  \href{http://arxiv.org/abs/hep-ph/9901246}{{\tt hep-ph/9901246}}.

\bibitem{Cahill-Rowley:2014ora}
M.~Cahill-Rowley, J.~Gainer, J.~Hewett, and T.~Rizzo, {\it {Towards a
  Supersymmetric Description of the Fermi Galactic Center Excess}},  {\em JHEP}
  {\bf 1502} (2015) 057, [\href{http://arxiv.org/abs/1409.1573}{{\tt
  arXiv:1409.1573}}].

\bibitem{Agrawal:2014oha}
P.~Agrawal, B.~Batell, P.~J. Fox, and R.~Harnik, {\it {WIMPs at the Galactic
  Center}},  \href{http://arxiv.org/abs/1411.2592}{{\tt arXiv:1411.2592}}.

\bibitem{Caron:2015wda}
A.~Achterberg, S.~Caron, L.~Hendriks, R.~Ruiz~de Austri, and C.~Weniger, {\it
  {A description of the Galactic Center excess in the Minimal Supersymmetric
  Standard Model}},  \href{http://arxiv.org/abs/1502.05703}{{\tt
  arXiv:1502.05703}}.

\bibitem{ATLAS:2014wva}
{\bf ATLAS, CDF, CMS, D0} Collaboration, {\it {First combination of Tevatron
  and LHC measurements of the top-quark mass}},
  \href{http://arxiv.org/abs/1403.4427}{{\tt arXiv:1403.4427}}.

\bibitem{Pato:2010zk}
M.~Pato, L.~Baudis, G.~Bertone, R.~Ruiz~de Austri, L.~E. Strigari, et~al., {\it
  {Complementarity of Dark Matter Direct Detection Targets}},  {\em Phys.Rev.}
  {\bf D83} (2011) 083505, [\href{http://arxiv.org/abs/1012.3458}{{\tt
  arXiv:1012.3458}}].

\bibitem{Bertone:2010rv}
G.~Bertone, D.~G. Cerdeno, M.~Fornasa, R.~R. de~Austri, and R.~Trotta, {\it
  {Identification of Dark Matter particles with LHC and direct detection
  data}},  {\em Phys.Rev.} {\bf D82} (2010) 055008,
  [\href{http://arxiv.org/abs/1005.4280}{{\tt arXiv:1005.4280}}].

\bibitem{ALEPH:2005ab}
{\bf ALEPH, DELPHI, L3, OPAL, SLD, LEP Electroweak Working Group, SLD
  Electroweak Group, SLD Heavy Flavour Group} Collaboration, S.~Schael et~al.,
  {\it {Precision electroweak measurements on the $Z$ resonance}},  {\em
  Phys.Rept.} {\bf 427} (2006) 257--454,
  [\href{http://arxiv.org/abs/hep-ex/0509008}{{\tt hep-ex/0509008}}].

\bibitem{Strege:2014ija}
C.~Strege, G.~Bertone, G.~Besjes, S.~Caron, R.~Ruiz~de Austri, et~al., {\it
  {Profile likelihood maps of a 15-dimensional MSSM}},  {\em JHEP} {\bf 1409}
  (2014) 081, [\href{http://arxiv.org/abs/1405.0622}{{\tt arXiv:1405.0622}}].

\bibitem{Arbey:2012ax}
A.~Arbey, M.~Battaglia, F.~Mahmoudi, and D.~Martínez~Santos, {\it
  {Supersymmetry confronts $B_s → μ^+μ^-$ : Present and future status}},
  {\em Phys.Rev.} {\bf D87} (2013), no.~3 035026,
  [\href{http://arxiv.org/abs/1212.4887}{{\tt arXiv:1212.4887}}].

\bibitem{CMSandLHCbCollaborations:2013pla}
{\bf CMS, LHCb} Collaboration, CMS and L.~Collaborations, {\it {Combination of
  results on the rare decays $B^0_{(s)} \to \mu^+\mu^-$ from the CMS and LHCb
  experiments}}, .

\bibitem{Bertone:2010ww}
G.~Bertone, K.~Kong, R.~R. de~Austri, and R.~Trotta, {\it {Global fits of the
  Minimal Universal Extra Dimensions scenario}},  {\em Phys.Rev.} {\bf D83}
  (2011) 036008, [\href{http://arxiv.org/abs/1010.2023}{{\tt
  arXiv:1010.2023}}].

\bibitem{Ade:2013zuv}
{\bf Planck} Collaboration, P.~Ade et~al., {\it {Planck 2013 results. XVI.
  Cosmological parameters}},  {\em Astron.Astrophys.} {\bf 571} (2014) A16,
  [\href{http://arxiv.org/abs/1303.5076}{{\tt arXiv:1303.5076}}].

\bibitem{Akerib:2013tjd}
{\bf LUX} Collaboration, D.~Akerib et~al., {\it {First results from the LUX
  dark matter experiment at the Sanford Underground Research Facility}},  {\em
  Phys.Rev.Lett.} {\bf 112} (2014) 091303,
  [\href{http://arxiv.org/abs/1310.8214}{{\tt arXiv:1310.8214}}].

\bibitem{Savage:2015xta}
C.~Savage, A.~Scaffidi, M.~White, and A.~G. Williams, {\it {LUX likelihood and
  limits on spin-independent and spin-dependent WIMP couplings with LUXCalc}},
  \href{http://arxiv.org/abs/1502.02667}{{\tt arXiv:1502.02667}}.

\bibitem{QCDSF:2011aa}
{\bf QCDSF} Collaboration, G.~S. Bali et~al., {\it {Strangeness Contribution to
  the Proton Spin from Lattice QCD}},  {\em Phys.Rev.Lett.} {\bf 108} (2012)
  222001, [\href{http://arxiv.org/abs/1112.3354}{{\tt arXiv:1112.3354}}].

\bibitem{Junnarkar:2013ac}
P.~Junnarkar and A.~Walker-Loud, {\it {Scalar strange content of the nucleon
  from lattice QCD}},  {\em Phys.Rev.} {\bf D87} (2013) 114510,
  [\href{http://arxiv.org/abs/1301.1114}{{\tt arXiv:1301.1114}}].

\bibitem{Aartsen:2012kia}
{\bf IceCube} Collaboration, M.~Aartsen et~al., {\it {Search for dark matter
  annihilations in the Sun with the 79-string IceCube detector}},  {\em
  Phys.Rev.Lett.} {\bf 110} (2013), no.~13 131302,
  [\href{http://arxiv.org/abs/1212.4097}{{\tt arXiv:1212.4097}}].

\bibitem{PhysRevLett.114.141301}
{\bf Super-Kamiokande} Collaboration, K.~Choi et~al., {\it Search for neutrinos
  from annihilation of captured low-mass dark matter particles in the sun by
  super-kamiokande},  {\em Phys. Rev. Lett.} {\bf 114} (Apr, 2015) 141301.

\bibitem{Adrian-Martinez:2013ayv}
{\bf ANTARES} Collaboration, S.~Adrian-Martinez et~al., {\it {First results on
  dark matter annihilation in the Sun using the ANTARES neutrino telescope}},
  {\em JCAP} {\bf 1311} (2013) 032, [\href{http://arxiv.org/abs/1302.6516}{{\tt
  arXiv:1302.6516}}].

\bibitem{deAustri:2006pe}
R.~R. de~Austri, R.~Trotta, and L.~Roszkowski, {\it {A Markov chain Monte Carlo
  analysis of the CMSSM}},  {\em JHEP} {\bf 0605} (2006) 002,
  [\href{http://arxiv.org/abs/hep-ph/0602028}{{\tt hep-ph/0602028}}].

\bibitem{Bechtle:2013wla}
P.~Bechtle, O.~Brein, S.~Heinemeyer, O.~Stål, T.~Stefaniak, et~al., {\it
  {$\mathsf{HiggsBounds}-4$: Improved Tests of Extended Higgs Sectors against
  Exclusion Bounds from LEP, the Tevatron and the LHC}},  {\em Eur.Phys.J.}
  {\bf C74} (2014), no.~3 2693, [\href{http://arxiv.org/abs/1311.0055}{{\tt
  arXiv:1311.0055}}].

\bibitem{Aad:2012tfa}
{\bf ATLAS} Collaboration, G.~Aad et~al., {\it {Observation of a new particle
  in the search for the Standard Model Higgs boson with the ATLAS detector at
  the LHC}},  {\em Phys.Lett.} {\bf B716} (2012) 1--29,
  [\href{http://arxiv.org/abs/1207.7214}{{\tt arXiv:1207.7214}}].

\bibitem{Chatrchyan:2012ufa}
{\bf CMS} Collaboration, S.~Chatrchyan et~al., {\it {Observation of a new boson
  at a mass of 125 GeV with the CMS experiment at the LHC}},  {\em Phys.Lett.}
  {\bf B716} (2012) 30--61, [\href{http://arxiv.org/abs/1207.7235}{{\tt
  arXiv:1207.7235}}].

\bibitem{Bechtle:2013xfa}
P.~Bechtle, S.~Heinemeyer, O.~Stål, T.~Stefaniak, and G.~Weiglein, {\it
  {$HiggsSignals$: Confronting arbitrary Higgs sectors with measurements at the
  Tevatron and the LHC}},  {\em Eur.Phys.J.} {\bf C74} (2014), no.~2 2711,
  [\href{http://arxiv.org/abs/1305.1933}{{\tt arXiv:1305.1933}}].

\bibitem{Djouadi:2006bz}
A.~Djouadi, M.~Muhlleitner, and M.~Spira, {\it {Decays of supersymmetric
  particles: The Program SUSY-HIT (SUspect-SdecaY-Hdecay-InTerface)}},  {\em
  Acta Phys.Polon.} {\bf B38} (2007) 635--644,
  [\href{http://arxiv.org/abs/hep-ph/0609292}{{\tt hep-ph/0609292}}].

\bibitem{Sjostrand:2014zea}
T.~Sjöstrand, S.~Ask, J.~R. Christiansen, R.~Corke, N.~Desai, et~al., {\it {An
  Introduction to PYTHIA 8.2}},  {\em Comput.Phys.Commun.} {\bf 191} (2015)
  159--177, [\href{http://arxiv.org/abs/1410.3012}{{\tt arXiv:1410.3012}}].

\bibitem{Ball:2012cx}
R.~D. Ball, V.~Bertone, S.~Carrazza, C.~S. Deans, L.~Del~Debbio, et~al., {\it
  {Parton distributions with LHC data}},  {\em Nucl.Phys.} {\bf B867} (2013)
  244--289, [\href{http://arxiv.org/abs/1207.1303}{{\tt arXiv:1207.1303}}].

\bibitem{Drees:2013wra}
M.~Drees, H.~Dreiner, D.~Schmeier, J.~Tattersall, and J.~S. Kim, {\it
  {CheckMATE: Confronting your Favourite New Physics Model with LHC Data}},
  {\em Comput.Phys.Commun.} {\bf 187} (2014) 227--265,
  [\href{http://arxiv.org/abs/1312.2591}{{\tt arXiv:1312.2591}}].

\bibitem{deFavereau:2013fsa}
{\bf DELPHES 3} Collaboration, J.~de~Favereau et~al., {\it {DELPHES 3, A
  modular framework for fast simulation of a generic collider experiment}},
  {\em JHEP} {\bf 1402} (2014) 057, [\href{http://arxiv.org/abs/1307.6346}{{\tt
  arXiv:1307.6346}}].

\bibitem{Ackermann:2015zua}
{\bf Fermi-LAT} Collaboration, M.~Ackermann et~al., {\it {Searching for Dark
  Matter Annihilation from Milky Way Dwarf Spheroidal Galaxies with Six Years
  of Fermi-LAT Data}},  \href{http://arxiv.org/abs/1503.02641}{{\tt
  arXiv:1503.02641}}.

\bibitem{Feroz:2008xx}
F.~Feroz, M.~Hobson, and M.~Bridges, {\it {MultiNest: an efficient and robust
  Bayesian inference tool for cosmology and particle physics}},  {\em
  Mon.Not.Roy.Astron.Soc.} {\bf 398} (2009) 1601--1614,
  [\href{http://arxiv.org/abs/0809.3437}{{\tt arXiv:0809.3437}}].

\bibitem{2015arXiv150608309S}
M.~{Schumann}, L.~{Baudis}, L.~{B{\"u}tikofer}, A.~{Kish}, and M.~{Selvi}, {\it
  {Dark matter sensitivity of multi-ton liquid xenon detectors}},  {\em ArXiv
  e-prints} (June, 2015) [\href{http://arxiv.org/abs/1506.08309}{{\tt
  arXiv:1506.08309}}].

\bibitem{Drees:1993bu}
M.~Drees and M.~Nojiri, {\it {Neutralino - nucleon scattering revisited}},
  {\em Phys. Rev.} {\bf D48} (1993) 3483--3501,
  [\href{http://arxiv.org/abs/hep-ph/9307208}{{\tt hep-ph/9307208}}].

\bibitem{Mandic:2000jz}
V.~Mandic, A.~Pierce, P.~Gondolo, and H.~Murayama, {\it {The Lower bound on the
  neutralino nucleon cross-section}},
  \href{http://arxiv.org/abs/hep-ph/0008022}{{\tt hep-ph/0008022}}.

\bibitem{Billard:2014yka}
J.~Billard, L.~Strigari, and E.~Figueroa-Feliciano, {\it {Solar neutrino
  physics with low-threshold dark matter detectors}},  {\em Phys. Rev.} {\bf
  D91} (2015), no.~9 095023, [\href{http://arxiv.org/abs/1409.0050}{{\tt
  arXiv:1409.0050}}].

\bibitem{Garny:2012it}
M.~Garny, A.~Ibarra, M.~Pato, and S.~Vogl, {\it {On the spin-dependent
  sensitivity of XENON100}},  {\em Phys.Rev.} {\bf D87} (2013), no.~5 056002,
  [\href{http://arxiv.org/abs/1211.4573}{{\tt arXiv:1211.4573}}].

\bibitem{Scott:2012mq}
{\bf IceCube} Collaboration, P.~Scott et~al., {\it {Use of event-level neutrino
  telescope data in global fits for theories of new physics}},  {\em JCAP} {\bf
  1211} (2012) 057, [\href{http://arxiv.org/abs/1207.0810}{{\tt
  arXiv:1207.0810}}].

\bibitem{IC79future}
{\bf IceCube} Collaboration, I.~Collaboration, {\it {}},  {\em to appear}
  (2015).

\bibitem{Bringmann:2012ez}
T.~Bringmann and C.~Weniger, {\it {Gamma Ray Signals from Dark Matter:
  Concepts, Status and Prospects}},  {\em Phys. Dark Univ.} {\bf 1} (2012)
  194--217, [\href{http://arxiv.org/abs/1208.5481}{{\tt arXiv:1208.5481}}].

\bibitem{Aad:2014kra}
{\bf ATLAS} Collaboration, G.~Aad et~al., {\it {Search for top squark pair
  production in final states with one isolated lepton, jets, and missing
  transverse momentum in $\sqrt s =$8 TeV $pp$ collisions with the ATLAS
  detector}},  {\em JHEP} {\bf 1411} (2014) 118,
  [\href{http://arxiv.org/abs/1407.0583}{{\tt arXiv:1407.0583}}].

\end{thebibliography}\endgroup

\end{document}